\documentclass{ws-ijmpd}
\usepackage[super,compress]{cite}
\usepackage{graphicx}
\usepackage{color}
\usepackage{multirow}
\usepackage[breaklinks]{hyperref}
\hypersetup{colorlinks,urlcolor=black,citecolor=black,linkcolor=black,filecolor=black}
\usepackage{breakurl}

\def \jcap {J. Cosmology Astropart. Phys.} 
\def \ijmpd {Int.\ J.\ Mod.\ Phys.\ D}

\def \araa {ARA\&A}
\def \aj {AJ}

\def \apj {ApJ}

\def \mnras {Mon.\ Non.\ Roy.\ Astron.\ Soc.}

\def \prd {Phys.\ Rev.\ D.}
\def \prl {Phys.\ Rev.\ Lett.}
\def\apj{Ap.\ J.}  
\def\mnras{MNRAS}

\def\prd{Phys.~Rev. {\bf D}}

\def\jcap{JCAP}

\begin{document}

\markboth{A. Bonilla \& J.E. Garc\'ia-Farieta}
{Exploring the Dark Universe}

%
\catchline{}{}{}{}{}
%

\title{Exploring the Dark Universe: constraints on dynamical Dark Energy models from CMB, BAO and growth rate measurements}

\author{ALEXANDER BONILLA RIVERA}

\address{Departamento de F\'isica, Universidade Federal de Juiz de Fora, 36036-330, Juiz de Fora, MG, Brazil.\\
abonilla@fisica.ufjf.br}

\author{JORGE ENRIQUE GARC\'IA-FARIETA}

\address{Departamento de F\'isica, Universidad Nacional de Colombia - Sede Bogot\'a, Carrera 45 No. 26-85, Bogot\'a, Colombia.\\
joegarciafa@unal.edu.co}

\maketitle

\begin{history}
\received{Day Month Year}
\revised{Day Month Year}
\end{history}

\begin{abstract}
In order to explain the current acceleration of the Universe, the fine tuning problem of the cosmological constant $\Lambda$ and the cosmic coincidence problem, different alternative models have been proposed in the literature. We use the most recent observational data from CMB (Planck 2018 final data release) and LSS (SDSS, WiggleZ, VIPERS) to constrain dynamical dark energy (DE) models. The CMB shift parameter, which traditionally has been used to determine the main cosmological parameters of the standard model $\Lambda CDM$ is employed in addition to data from redshift-space distortions through the growth parameter $A(z)=f(z)\sigma_{8}(z)$ to constrain the mass variance $\sigma_{8}$. BAO data is also used to study the history of the cosmological expansion and the main properties of DE. From the evolution of $q(z)$ we found a slowdown of acceleration behaviour at low redshifts, and by using the Akaike and Bayesian Information Criterion (AIC, BIC) we discriminate different models those that are better suited to the observational data, finding that the interactive dark energy (IDE) model is the most favoured by observational data, including information from SNIa and Hz. The analysis shows that the IDE model is followed closely by EDE and $\Lambda CDM$ models, which in some cases fit better the observational data with individual probes.
\end{abstract}


\ccode{PACS numbers: 95.35.+d, 95.36.+x, 98.80.-k, 98.80.Es}

\section{Introduction}\label{sec:intro}

In the last few decades the $\Lambda$CDM model has been the most famous 
cosmological model, which together with the inflationary paradigm, 
predicts the hierarchical structure formation with a total composition 
around $4\%$ baryons, $26\%$ dark matter (DM) and the remaining $70\%$ 
of a non very well known component called dark energy (DE), which is 
traditionally presented as the main responsible of the late accelerated 
expansion of the Universe. In the $\Lambda$CDM model, DM is composed by collisionless non baryonic particles and DE is described in terms of the cosmological constant $\Lambda$ with an equation of state (EoS) $w=-1$. The predictions made by this model are in very good agreement with observations of the anisotropies in the Cosmic Microwave Background (CMB) radiation, baryonic acoustic oscillations (BAO), Supernovae Ia (SNIa), etc. Nevertheless, the standard model has fundamental problems related to the nature of DM and DE \cite{2008ARA&A..46..385F,2006astro.ph..9591A}. In the context of DE, there are several theoretical arguments against a cosmological constant. One of them is the coincidence problem, associated to the order of magnitude of DE and DM densities at present epoch, i.e., $\Omega_m \approx \Omega_{\Lambda}$. An additional issue is related to the fine tuning of the current value of $\Lambda$, being quite far from high energy particle physics predictions \cite{weinberg89,cop06}. In order to alleviate these issues caused by the introduction of a cosmological constant, several DE models with dynamical EoS have been proposed \cite{2008ARA&A..46..385F}. For instance, the EoS can be parametrized in different ways, being one of the most popular the Chevallier-Polarski-Linder (CPL) \cite{Chevallier:2000qy, Linder:2003nc}. Other models, instead, include a scalar field which mimic the role of DE, for example, quintessence \cite{Caldwell:1998, Ratra:1988}, phantom \cite{cal02, chi00,par99}, quintom \cite{Guo:2004} and k-essence fields \cite{armendariza,armendarizb, chi00}. In addition, there are several DE models that consider interactions with DM in order to solve the cosmic coincidence problem, e.g. Interacting Dark Energy (IDE) \cite{CalderaCabral:2008bx}, the Holographic DE (HolDE) \cite{coh99,suss95,hooft93, hooft01,zhang10}, modified gravity (ModGrav)\cite{Dvali:2000hr} and Braneworld models \cite{garcia}, are free from the cosmological constant problem. In general these families of models can be classified as phantom if EoS $\omega < -1$, or as quintessence if $\omega > -1$; in the first case a fluid multicomponent is required with at least one phantom constituent, which has been shown to suffer serious theoretical problems, and in the second case, general relativity needs to be extended to a more general theory at cosmological scales \cite{Nesseris:2006er}. Among these families of models we have to discriminate which one is the most favoured by current observations. The most popular method of discrimination is through the Akaike and Bayesian Information Criteria \cite{Shi:2012ma,Akaike:74,Schwarz:78}, which indicate which model fits better the observational data taking into account the number of free parameters and data points of each model. In order to compare models with observations we use data from SNIa, CMB, BAO,.., etc, which are considered as \textit{geometric test} and allow us to determine $H(z)$ independently of the Einstein's equations validity, directly through the redshift dependence with cosmological distances (e.g. the angular diameter distance $d_A (z)$ and mass gas fraction $f_{gas}$, among others). A different approach to determine $H(z)$ is by implementing \textit{dynamical tests}, that measure the evolution of the density field (background or perturbations) connecting it with the geometry through a theory of gravity. An example of a dynamical test of geometry is given by measuring the linear growth factor of matter density perturbations $D(a)$, whose value can be obtained by different methods like redshift distortion factor from redshift surveys [$A(z)=f(z)\sigma_{8}(z)$], number counts of galaxy clusters ($dN(M,z)/dMdz$), large-scale structure power spectrum [$P(k)$] and Integrated Sachs-Wolfe (ISW) effect. In this paper we include this test through data from $ A_{obs} (z) $, which is important to understand the effects of DE on the growth of structures.\\

Our aim in this paper is to constrain the main set of parameters in some of the well established models of DE by using CMB, BAO and growth rate of LSS observational test, in the frame of the Friedmann-Lema\^itre-Robertson-Walker (FLRW) cosmology. The paper is organized as follows. In $\oint 2$ we describe the cosmological tests and the datasets used, in $\oint 3$, we introduce the statistical tools to discriminated models given the Bayesian analysis performed. The cosmological models and their analysis are presented in $\oint 4$, and a study of the history of expansion through the deceleration parameter is given in $\oint 5$. Finally, in section $\oint 6$ we conclude with a summary and discussion of our results.

\section{The cosmological tests}

Fluctuations in the density field make evolve each component of the cosmic fluid (DM, baryonic matter and photons), at different ratios by their interactions in a gravitational potential \cite{2008CamUnivPres,1989KolbTurner}. These fluctuations grow through gravitational instabilities as the Universe expands, as consequence the matter and radiation decouple creating the CMB radiation and in matter domain the Large Scale Structure (LSS) of the Universe is formed. In early stages of the Universe the radiation was energetic enough to ionize hydrogen, consequently the interactions by Thomson scattering in the radiation coupled to baryons forms a photon-baryon fluid. Taking into account that the radiation pressure due to photons is opposed to the gravitational compression of the fluid, the fluctuations in density produces a harmonic motion whose amplitude does not grow but slowly decays originating the so-called Baryon Acoustic Oscillations (BAO). These patterns printed in the density field can be observed in the power spectrum of matter and radiation \cite{2005APh....24..334W}. Subsequently the photons are diluted with the cosmic expansion and stream out of potential wells. Although effectively without pressure, the baryons still contribute to the inertia and gravitational mass of the fluid, producing changes in the balance of pressure and gravity, the resulting effect is that baryons drag photons to potential wells. After these processes the perturbations in the photon-baryon fluid propagate as acoustic waves with sound speed $c_s$, defining a sound comoving horizon $r_s$ at the epoch of drag ($z_{drag}$). In the recombination epoch ($z_{cmb}$), photons are decoupled from matter, and baryons can now constitute neutral elements while radiation is scattered for last time, forming the so-called CMB \cite{2008CamUnivPres} mentioned above. The resultant fluctuations in CMB observed in radiation maps, with anisotropy order around $\Delta T/T \sim 10^{-5}$ are better studied with its power spectrum.\\

In the following sections we present the details of the observational samples used to perform the Bayesian analysis: CMB by using Shift parameter $R$; BAO by means of Distance Ratio Scale $D_v (z)/r_s$ and growth rate of LSS through Growth Parameter $A(z)=f(z)\sigma_{8}(z)$, adopted in order to constrain the free parameters for each cosmological model considered, including some derived parameters as Table \ref{tab:par1} shows.\\

\begin{table*}
\tbl{Notation and short overview of the cosmological parameters used in this analysis. The upper block contains the main set of free parameters used in the Bayesian analysis. The lower block displays the derived parameters for each model.}
{\begin{tabular}{ll}
\hline 
\hline 
Parameter            & Physical meaning  and/or definition    \\ 
\hline 
\hline 
\textit{h} &  Dimensionless Hubble parameter \\ 
$\Omega_m$   & Dimensionless DM density parameter\\ 
$\Omega_\Lambda$       &  Dimensionless DE density parameter to $\Lambda CDM$\\ 
$\Omega_k$         &   Dimensionless curvature density parameter \\ 
$\sigma_8$        &    RMS matter fluctuations at 8Mpc/h in linear theory \\ 
$\omega$          &   Constant EoS to $\omega CDM$ \\ 
$\omega (a) = \omega_0 + (1+a)\omega_1$ &  EoS for \textit{CPL} parametrization  \\
$\omega_x$, $\delta$         &   EoS and dimensionless coupling term for \textit{IDE} \\ 
$\omega_0$, $\Omega_e$         &   EoS and asymptotic DE density term for \textit{EDE}\\ 
\hline 
$H_0 = 100 h$ &   Current expansion rate (Hubble parameter) in $Km.s^{-1} Mpc^{-1}$\\ 
$t_0$ &   Age of the Universe today (in Gyr) \\ 
$\Omega_b=0.045$   &  Dimensionless baryon density parameter \\ 
$\Omega_r =\Omega_\gamma + \Omega_\nu$   &   Dimensionless radiation density parameter \\ 
$\Omega_\gamma =2.469 \times 10^{-5} h^{-2}$   &  Dimensionless photon density parameter\\
$\Omega_\nu$   &  Dimensionless neutrino density parameter \\
$N_{eff}=3.04$   &  Effective number of relativistic neutrino degrees of freedom\\
$\omega_m=\Omega_m h^2$   &   Physical DM density\\ 
$\omega_b=\Omega_b h^2$   &   Physical  baryon density\\ 
$\rho_{cri}= 3 H_0^2 / 8\pi G$   & Critical density ($1.88 \times10^{29} h^2g/cm^3$) \\ 
$\Omega_X$   &  Dimensionless DE density parameter\\ 
$\rho_X= \rho_{cri} \Omega_X$  &  Physical  DE density   \\
$\Lambda =  8\pi G \rho_{\Lambda}$  &  Cosmological constant where  $\rho_{\Lambda} = \rho_{cri} 3 H_0^2$ \\
$c_s$  &  Sound speed  \\ 
$r_s$  &  Comoving size of sound horizon  \\  
$z_{drag}$ &  Redshift at which baryon-drag optical depth equals unity \\ 
$r_{drag} = r_s (z_{drag})$ &  Comoving size of the sound horizon at $z_{drag}$\\
$r_{s}/D_v (z)$ & BAO distance ratio scale \\
$z_{cmb}$ &  Redshift at decoupled photon-baryon  \\
$R(z_{cmb})$ & Scaled distance at recombination ($z_{cmb}$) \\
$l_A (z_{cmb})$ & Angular scale of sound horizon at recombination ($z_{cmb}$) \\
\hline \hline
\end{tabular}\label{tab:par1}}
\end{table*}

\subsection{CMB}

To explore the expansion history in each model, we use CMB information from Planck 2018 data \cite{2018arXiv180706209P}. A particular test to probe DE is given by the angular scale of sound horizon $r_s$, at decoupling time ($z_{cmb}\sim 1090$), which is encrypted in the $l^{TT}_{1}$ mode of the first peak of the CMB power spectrum. The $\chi^2$ for the CMB data is constructed as
\begin{equation}
 \chi^2_{CMB} = X_{Planck18}^TC_{cmb}^{-1}X_{Planck18},
 \label{eq3:3.1}
\end{equation}
\noindent such that
\begin{equation}
 X _{Planck18}=\left(
 \begin{array}{c}
  R - 1.7502 \\
 l_A - 301.471 \\
\omega_b - 0.02236
\end{array}\right),
 \label{eq3:3.2}
\end{equation}
\noindent where $\omega_b = \Omega_b h^2$ \cite{2015JCAP...12..022H}. Here $l_A$ is the acoustic scale defined as
\begin{equation}
l_A = \frac{\pi d_A(z_{cmb})(1+z_{cmb})}{r_s(z_{cmb})},
 \label{eq3:3.3}
\end{equation}
\noindent with $d_A(z_{cmb})$ being the angular diameter distance and $z_{cmb}$ the redshift of decoupling given by \cite{husugi},
\begin{equation}
z_{cmb} = 1048[1+0.00124(\Omega_b h^2)^{-0.738}]
[1+g_1(\Omega_{m}h^2)^{g_2}],
 \label{eq3:3.4}
\end{equation}
\begin{equation}
g_1 = \frac{0.0783(\Omega_b h^2)^{-0.238}}{1+39.5(\Omega_b
h^2)^{0.763}},~~ g_2 = \frac{0.560}{1+21.1(\Omega_b h^2)^{1.81}}.
  \label{eq3:3.5}
\end{equation}
The shift parameter $R$ is defined as \cite{BET97}
\begin{equation}
R = \frac{\sqrt{\Omega_{m}}}{c} d_A (z_{cmb}) (1+z_{cmb}).
 \label{eq3:3.6}
\end{equation}
The term $C_{cmb}^{-1}$ in Eq. (\ref{eq3:3.1}) corresponds to the inverse covariance matrix for ($R, l_A, \omega_b$), that with Planck 2018 data is equivalent to $C_{cmb^{Planck18}}^{-1} = \sigma_{i} \sigma_{j} C_{NorCov_{i,j}}$, with $\sigma_{i} = \left( 0.0046, 0.090, 0.00015 \right)$, in which case this test contributes with 3 data points to the statistical analysis, considering that the full normalised covariance matrix \cite{2018arXiv180805724C} is given by
\begin{equation}
C_{NorCov_{i,j}} = \left(
\begin{array}{ccc}
1.00 & 0.46 & -0.66\\
0.46 &  1.00 & -0.37\\
-0.66 & -0.33 & 1.00
\end{array}\right).
 \label{eq3:3.10}
\end{equation}

\subsection{BAO}
The large scale correlation function measured from the 2dF Galaxy Redshift Survey and SDSS redshift survey, displays a peak around $150 h^{-1}Mpc$ in comoving coordinates \cite{2005MNRAS.362..505C,2005ApJ...633..560E}, which is related to the expanding spherical wave of baryonic perturbations from acoustic oscillations at recombination time. As previously mentioned, BAO correspond to periodic fluctuations in the density field, printed in the primordial plasma before decoupling, that can be used as standard rule to characterize the properties of DE \cite{2010deot.book..246B,2005APh....24..334W}. To obtain constraints on a certain cosmological model we consider the $\chi^2$ for WiggleZ BAO data \cite{2011MNRAS.415.2892B} given by 
\begin{equation}
\chi^2_{\scriptscriptstyle WiggleZ} =
(\bar{A}_{obs}-\bar{A}_{th})C_{\scriptscriptstyle
WiggleZ}^{-1}(\bar{A}_{obs}-\bar{A}_{th})^T,
\end{equation}
\noindent where $\bar{A}_{obs} = (0.447, 0.442, 0.424)$ is the data vector at $z=(0.44,0.60,0.73)$ and $\bar{A}_{th}(z,p_i)$ is the theoretical predicted value 
given by \cite{2005ApJ...633..560E}
\begin{equation}
\bar{A}_{th}=D_V(z) \frac{\sqrt{\Omega_m H_0^2}}{cz},
\end{equation}
\noindent assuming the distance scale $D_V(z)$ defined traditionally as
\begin{equation}
D_V(z) = \frac{1}{H_0}\left[ (1+z)^2 d_A (z)^2
\frac{cz}{E(z)}\right]^{1/3},
 \end{equation}
\noindent with $d_A(z)$ being the angular diameter distance. Additionally, the inverse covariance matrix for the WiggleZ dataset $C_{\scriptscriptstyle WiggleZ}^{-1}$ can be expressed explicitly as
\begin{equation}
C_{\scriptscriptstyle WiggleZ}^{-1} = \left(
\begin{array}{ccc}
1040.3 & -807.5   & 336.8    \\
-807.5  & 3720.3  & -1551.9 \\
336.8   & -1551.9 & 2914.9
\end{array}\right).
\end{equation}

\begin{figure}[htb]
\centering
\begin{center}
    \includegraphics[height=8cm]{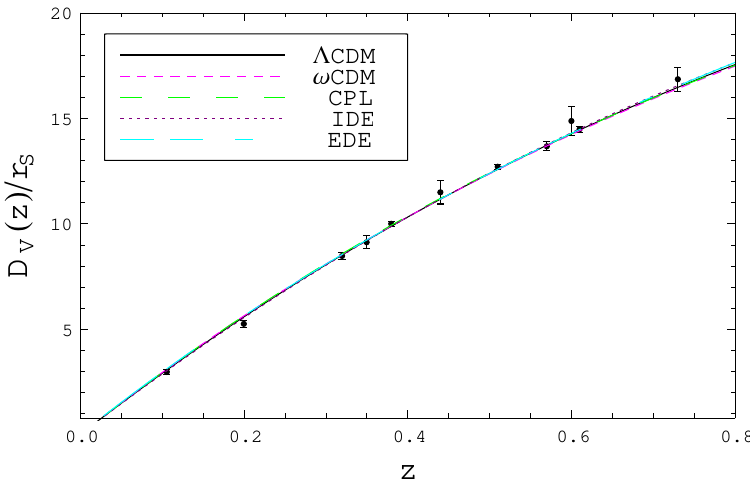}
    \caption{The distance-redshift relation using best-fit values for BAO+CMB+G and BAO measurements $D_v (z)/r_s$ for each model considered in this paper.}
    \label{fig:Dvz}
\end{center}
\end{figure}

Similarly, for the SDSS DR7 - BAO distance measurements, the $\chi^2$ can be expressed as \cite{2010MNRAS.401.2148P}
\begin{equation}
\chi^2_{\scriptscriptstyle SDSS} =
(\bar{d}_{obs}-\bar{d}_{th})C_{\scriptscriptstyle
SDSS}^{-1}(\bar{d}_{obs}-\bar{d}_{th})^T,
 \label{eq3:3.19}
\end{equation}
\noindent where $\bar{d}_{obs} = (0.1905,0.1097)$ is measured at $z=0.2$ and $z=0.35$, whereas $\bar{d}_{th}(z_d,p_i)$
denotes the distance ratio
\begin{equation}
\bar{d}_{th} = \frac{r_s(z_d)}{D_V(z)},
 \label{eq3:3.20}
\end{equation}
\noindent where $r_s(z)$ is the comoving sound horizon
given by
\begin{equation}
 r_s(z) = c \int_z^\infty \frac{c_s(z')}{H(z')}dz',
  \label{eq3:3.13}
 \end{equation}
\noindent and $c_s(z)$ is the sound speed
\begin{equation}
c_s(z) = \frac{1}{\sqrt{3(1+\bar{R_b}/(1+z))}},
 \label{eq3:3.14}
\end{equation}
\noindent with $\bar{R_b} = 31500\Omega_{b}h^2(T_{CMB}/2.7\rm{K})^{-4}$ and $T_{CMB} = 2.726K$. The redshift $z_{drag}$ at the baryon drag epoch is fitted with the formula \cite{1998ApJ...496..605E},
\begin{equation}
z_{drag} =
\frac{1291(\Omega_{m}h^2)^{0.251}}{1+0.659(\Omega_{m}h^2)^{0.828}}[1+b_1(\Omega_b
h^2)^{b_2}],
 \label{eq3:3.15}
\end{equation}
\noindent where $b_1 = 0.313(\Omega_{m}h^2)^{-0.419}[1+0.607(\Omega_{m}h^2)^{0.674}]$ and $b_2 = 0.238(\Omega_{m}h^2)^{0.223}$. In this case, the inverse of the covariance matrix for the SDSS dataset $C_{\scriptscriptstyle SDSS}^{-1}$ is given by
\begin{equation}
C_{\scriptscriptstyle SDSS}^{-1} = \left(
\begin{array}{cc}
30124 & -17227\\
-17227 & 86977
\end{array}\right).
 \label{eq3:3.21}
\end{equation}
For the 6dFGS - BAO data \cite{2011MNRAS.416.3017B}, there is only one data point at $z=0.106$, so that the $\chi^2$ is computed by
\begin{equation}
\chi^2_{\scriptscriptstyle 6dFGS} =
\left(\frac{d_z-0.336}{0.015}\right)^2.
 \label{eq3:3.22}
\end{equation}
Additionally, we include measures from the Main Galaxy
Sample of Data Release 7 of Sloan Digital Sky Survey (SDSS-MGS)
\cite{bao3} ($r_s/D_V(0.57)=0.0732 \pm 0.0012$), the LOWZ
and CMASS galaxy samples of the Baryon Oscillation Spectroscopic
Survey (BOSS) \cite{bao3} ($D_V/r_s(0.32)=8.47 \pm 0.17$), the distribution of the
LymanForest in BOSS (BOSS - $Ly_{\alpha}$) \cite{bao4} ($D_A/r_s(2.36)=10.08 \pm 0.4$) and BOSS DR12 galaxy sample ($D_V/r_s(0.38)=1477\pm16$, $D_V/r_s(0.51)=1877\pm19$, $D_V/r_s(0.61)=2140\pm22$) (Fig. \ref{fig:Dvz}). Therefore, the total measurements and their corresponding effective redshifts include 12 data point and whose minimization is given by

\begin{eqnarray}
\chi_{BAO}^{2} &=& \chi^{2}_{WiggleZ} + \chi^{2}_{SDSS} + \chi^{2}_{6dF} + \chi^{2}_{SDSS-MGS} + \chi^{2}_{BOSS-LOWZ} \nonumber \\
 &&  + \chi^{2}_{BOSS-Ly_{\alpha}}
\end{eqnarray}

\subsection{Growth Rate of LSS}

\begin{figure}[htb]
\centering
\begin{center}
    \includegraphics[height=7.5cm]{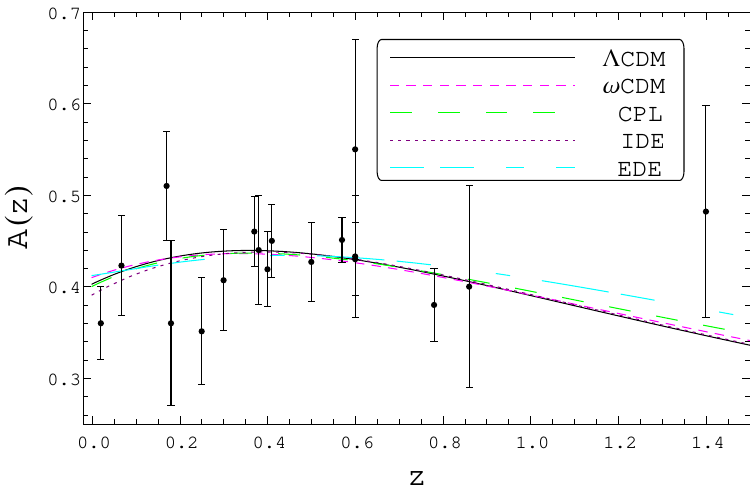}
    \caption{Growth rate measurements $A_{obs}(z_{i})$ and theoretical expectations for different cosmological models using bets fit values for BAO+CMB+G.}
    \label{fig:Az}
\end{center}
\end{figure}

The LSS of the Universe can be described geometrically in terms of vast empty regions, sheets, filaments, clusters of galaxies and superclusters. These structures evolved from a perturbed density field by gravitational collapse and can be treated theoretically, from a perturbation approach, as deviations from the mean density. Following the usual definition of the matter density contrast $\delta (r,t)\equiv \delta \rho (r,t) / \rho (r,t)$, the dynamics of the cosmic Hubble expansion $H(t)$ is driven by the gravitational field of the mean matter density $\rho (r,t)$, while the density fluctuations $\delta \rho (r,t)$ produces an additional gravitational field at first order of perturbation. In overdensed regions, $\delta \rho (r,t) > 0$, the gravitational field is stronger than the cosmic average and therefore, due to this excess the overdensed region will expand slower than the average. On the other hand, in underdensed regions, $\delta \rho (r,t) < 0$, the gravitational field is weaker than in the cosmic mean and therefore, the expansion is faster. Overdense regions increase their density contrast over time, while underdense regions decrease their density contrast, in both situations $\vert \delta \vert$ increase with time. The growth of the density perturbations can be characterised by assuming the following relationship $\delta (r,t) = D(t) \delta_0 (r)$, where $D(t)$ is the linear structure growth factor and $\delta_0 (r)$ is an arbitrary function of the spatial coordinates. Under the assumption that general relativity is the correct theory of gravity ($G_{eff}(a)=1$) \cite{2011JCAP...07..037N,2006astro.ph..5313U,2006PhRvD..74h4007S}, we characterize the growth of structures by using $D(a)$, obtained numerically from the following equation
\begin{equation}
\ddot{D}(a) + \left( \frac{3}{a} + \frac{\dot{H}(a)}{H(a)} \right) \dot{D}(a) - \frac{3}{2} \frac{\Omega_m}{a^5 H(a)^2} G_{eff}(a)D(a) = 0,
 \label{eq3:3.24}
\end{equation}
where dots denote differentiation with respect to the scale factor $a$ and initial conditions $D(0) = 0$ and $\dot{D}(0) = 1$ are assumed for the growing mode. The solution $\delta (r,t) = D(t) \delta_0 (r)$ indicates that in linear perturbation theory the spatial shape of the density fluctuations is frozen in comoving coordinates and only its amplitude increases. Besides, an observational estimate of the growth rate can be obtained from the linear growth factor through $f(a)\equiv a\dot{D}(a)/D(a)$, in which case we use the parameter $A(z)=f(z)\sigma_{8}(z)$ to constrain cosmological models by minimizing
\begin{equation}
\chi^{2}_{G} = \sum_{i=1}^n \frac{(A(z) - A_{obs}(z_i))^2}{\sigma_i^2},
 \label{eq3:3.25} 
\end{equation} 
where $\sigma_{8}(z)$ corresponds to the RMS mass fluctuation on spheres of $8Mpch^{-1}$ and $A_{obs}(z_i)$ is the observed growth parameter that includes the Alcock-Paczynski effect in redshift-space distortions (Fig. \ref{fig:Az}). The datasets used in this paper for the growth parameter were obtained from the following projects: PSCz, 2dF, VIPERS, SDSS, 2MASS, GAMA, WiggleZ and FastSound galaxy surveys (Table \ref{tab:Az}). Given $\sigma_{8}(z)=\sigma_{8}^0D(z)/D(0)$, we use $\sigma_{8}^0$ as a free parameter. To complement our analysis, we use 580 Supernovae data (SNIa) from Union2.1 \cite{Suzuki_2012} and 36 observational Hubble Data (OHD) from \cite{2015arXiv150702517M} (See appendix \ref{AppxA}).

\section{Method and data analysis}
We implemented a Bayesian analysis through maximizing the likelihood $\mathcal{L}$, this method allow to find the best-fit values for a certain set of parameters given a theoretical model. The maximum likelihood estimated for the best-fit parameters $\Theta_{i}^{m}$ is
\begin{equation}
 \mathcal{L}_{max}(\Theta_{i}^{m})= exp \left[ -\frac{1}{2}{\chi_{min}^{2}(\Theta_{i}^{m})} \right].
  \label{eq4:4.1}
\end{equation}
In our case $\mathcal{L}_{max}(\Theta_{i}^{m})$ has a Gaussian error distribution \cite{Andrae:2010gh}, so that the minimized $\chi^{2}$ distribution can be expressed as
\begin{equation}
\chi_{min}^{2}(\Theta_{i}^{m})=-2 \ln \mathcal{L}_{max}(\Theta_{i}^{m}).
\label{eq4:4.2} 
\end{equation}
Combining the different datasets and considering the properties of $\chi^{2}$, the final constrain of parameters is obtained from the full posterior distribution
\begin{equation}
\chi_{min}^{2}= \chi^{2}_{CMB} +\chi^{2}_{BAO}+\chi^{2}_{G}+\chi^{2}_{SNIa}+\chi^{2}_{H(z)}.
\label{eq4:4.3}
\end{equation}
To compute the uncertainties we use the Fisher matrix formalism, which is widely used in several analysis to constrain cosmological parameters from different observations \cite{Albrecht:2009ct,2012JCAP...09..009W}. The coefficients of the Fisher matrix encode the Gaussian uncertainties of the parameters $\Theta_{i}^{m}$ and they can be computed in terms of the best-fit $\chi_{min}^{2}$ as
\begin{equation}
F_{ij}=\frac{1}{2}\frac{\partial ^2 \chi_{min}^{2}}{\partial p_i \partial p_j},
\label{eq4:4.4} 
\end{equation}
where $p_i$ and $p_j$ are the set of free parameters in each model. In its extended form the Fisher matrix is given by 
\begin{equation}
 \left[ F \right] =  \frac{1}{2}\left[
\begin{array}{cccccc}
\frac{\partial^2}{\partial p_1^2}             & \frac{\partial^2}{\partial p_1 \partial p_2} & . & . & . & \frac{\partial^2}{\partial p_1 \partial p_n}\\
\frac{\partial^2}{\partial p_2 \partial p_1} & \frac{\partial^2}{\partial p_2^2} & . & . & . & \frac{\partial^2}{\partial p_2 \partial p_n}\\ 
.                                                                    & .                                                    & . &  &  & .\\ 
.                                                                    & .                                                    &  & . &  & .\\ 
.                                                                    & .                                                    &  &  & . & .\\ 
\frac{\partial^2}{\partial p_n \partial p_1} &  \frac{\partial^2}{\partial p_n \partial p_2} & . & . & . & \frac{\partial^2}{\partial p_n^2} \\ 
\end{array}\right] \chi_{min}^{2}(p_1, p_2,...,p_n ),
\label{eq4:4.5}
\end{equation}

with $\chi_{min}^{2}(p_1,p_2,...,p_n)=\chi_{min}^{2}(\Theta_{1}^{m},\Theta_{2}^{m},...,\Theta_{n}^{m})$ in this work. The inverse of the Fisher matrix corresponds to the covariance matrix $C_{cov}$ as given in Eq. (\ref{eq4:4.6}), where its coefficients $\sigma_{i}$ and $\sigma_{j}$ are the uncertainties associated to each parameter $p_i$ and $p_j$, with $1\sigma$ of statistical confidence. The uncertainties are obtained as $\sigma_i = \sqrt{Diag \left[ C_{cov} \right]_{ij} }$.
\begin{equation}
 \left[ C_{cov} \right] = \left[ F \right]^{-1} =  \left[
\begin{array}{cccccc}
\sigma^2_1  & \sigma_{12} & . & . & . & \sigma_{1n}\\
\sigma_{21} &  \sigma^2_2 & . & . & . & \sigma_{2n}\\ 
.                                                                    & .                                                    & . &  &  & .\\ 
.                                                                    & .                                                    &  & . &  & .\\ 
.                                                                    & .                                                    &  &  & . & .\\ 
\sigma_{n1} &  \sigma_{n2} & . & . & . & \sigma^2_{n}\\ 
\end{array}\right].
\label{eq4:4.6}
\end{equation}

As mentioned at the beginning, we are focused on obtaining tight constraints on the set of parameters in each cosmological model, to discern among them which model is the most favoured by the current observations. To accomplish this goal we compare the best-fit results using the Akaike information criterion (AIC), \cite{Akaike:74} and the Bayesian Information Criterion (BIC)\cite{Schwarz:78}, that allow to compare cosmological models with different degrees of freedom, with respect to the observational evidence and the set of parameters used \cite{Liddle:04}. The AIC and BIC can be computed as
\begin{equation}
AIC=-2 \ln \mathcal{L}_{max} +2k,
\label{eq4:4.7}
\end{equation}
\begin{equation}
BIC=-2 \ln \mathcal{L}_{max} + k \ln N,
\label{eq4:4.8} 
\end{equation}

\noindent where $\mathcal{L}_{max}$ is the maximum likelihood of the model under consideration, and $k$ is the number of parameters. Given the BIC criterion considers the number of data points $N$ used in the fit, it imposes a strict penalty against extra parameters for any set of data $\ln N > 2$. The preferred model corresponds to the one that minimizes AIC and BIC, 
for this reason we consider, instead of their absolute values, their relative values between the different models. Therefore the weight of the evidence can be characterised by $\Delta AIC=AIC_{i}-AIC_{min}$ and $\Delta BIC=BIC_{i}-BIC_{min}$, where the subindex $i$ refers to value of $AIC$ ($BIC$) for the model $i$ and $AIC_{min}$ ($BIC_{min}$) is the minimum value of $AIC$ ($BIC$) among all the models \cite{Burnham:03,Robert:95}. Tables \ref{tab:AIC} and \ref{tab:BIC} show the assignation adopted by each criterion in terms of their relative difference.
\begin{table*}[htb]
\tbl{$\Delta AIC$ criterion.}
{\begin{tabular}{cc}
\hline \hline
$\Delta AIC$ & Level of Empirical Support For Model $i$\\
\hline\hline
$0-2$ & Substantial \\
$4-7$ & Considerably Less  \\
$>10$ & Essentially None \\
\hline\hline
\end{tabular}\label{tab:AIC}}
\end{table*}
\begin{table*}[htb]
\tbl{$\Delta BIC$ criterion.}
{\begin{tabular}{cccc}
\hline
\hline
$\Delta BIC$ &  Evidence Against Model $i$\\
\hline\hline
$0-2$ & Not Worth More Than A Bare Mention  \\
$2-6$ & Positive   \\
$6-10$ & Strong  \\
$>10$ & Very Strong \\
\hline
\hline
\end{tabular}\label{tab:BIC}}
\end{table*}

To achieve the aims of this research we consider $N = 639$ data points from independent cosmological probes: CMB (3), BAO (12), G (18), SNIa (580), $H(z)$ (36). The priors used in the present analysis are standard and conservative as possible. Following the methodology exposed, in next section we present the main results obtained per model, and then a comparison using the $AIC$ and $BIC$, displays the hierarchy of the models preferred by the observations given their phenomenology related to DE.

\section{Cosmological models and results}\label{sec2}

In order to constrain DE models, we calculate the theoretical angular diameter distance predicted by a Friedmann-Lemaitre-Robertson-Walker (FLRW) metric, and compare it with the observations. For a source at redshift $z$, the angular diameter distance is given by
\begin{equation}
d_{A} (z,\Theta_i^m) =  \frac{3000h^{-1}}{(1+z)} \frac{1}{\sqrt{\mid \Omega_{k} \mid}} 
\sin \varsigma \left( \int_{0}^{z} \frac{\sqrt{\mid \Omega_{k} \mid}}{E(z,\Omega_i)}dz\right), 
\label{eq2:2.1}
\end{equation}
where $h$ is dimensionless Hubble parameter ($H_{0} = h 100 \mathrm{km}$ $\mathrm{s}^{-1}\mathrm{Mpc}^{-1}$) and the function $\sin \varsigma(x)$ is defined as $\sinh(x)$ if $\Omega_{k} >0$, $\sin(x)$ if $\Omega_{k} <0$ and $x$ if $\Omega_{k} =0$ \cite{Hogg:1999ad}. Currently, all the evidence of DE comes from measurements of the expansion rate $H(z)$ that provides a detail description for the expansion history of the Universe. In a standard FLRW cosmology, the expansion rate as a function of the redshift $H(z)$ is given by the Friedmann equation as
\begin{equation}
E^2(z,\Omega_i) = \Omega_r(1+z)^{4} + \Omega_m(1+z)^{3} + \Omega_k(1+z)^{2} + \Omega_X e^{3\int_0^z \frac{dz'}{1+z'} \left( 1+w(z')\right) }
\label{eq2:2.2}
\end{equation}

\noindent with $E(z,\Omega_i)=H(z)/H_0$, $H_0$ the Hubble parameter today, and the redshift relationship in terms of the scale factor $1+z=a^{-1}$. In the equation  (\ref{eq2:2.2}) $\Omega_i$ is the current energy density corresponding to radiation ($\Omega_{r}$), matter ($\Omega_{m}$), curvature ($\Omega_{k}$) and DE ($\Omega_{X}$), normalised respectively to the today's critical density $\rho_{cri}= 3 H_0^2 / 8\pi G$. The EoS of DE is characterised by the ratio pressure to energy-density $\omega (a)=p(a)/\rho (a)$, allowing to classify the models into two groups: one with energy density constant and the other with energy density dynamic. For each model, the density parameter of curvature $\Omega_{k}$ is free, and each one of them have a vector of parameters $\Theta_{i}^{model}= \left\lbrace \theta_i, \Omega_i \right\rbrace $, where $\theta_i = \left\lbrace h, \sigma_8 \right\rbrace$ and $\Omega_i = \left\lbrace  \Omega_r , \Omega_m , \Omega_k , \Omega_x  \right\rbrace$ is a multicomponent fluid for the analysis in this work.\\

The Hubble parameter $H(z)$ offers a natural description about the kinematics of the cosmic expansion and its dependence with time. In particular, to characterize whether the Universe is currently accelerating or decelerating, the history of expansion is fitted through the deceleration parameter $q(z)\equiv -\ddot{a}(z)/a(z)H(z)^2$. If $q(z)>0$, it means $\ddot{a}(z)<0$, then the expansion decelerate as expected due to the gravitational collapse. Despite the fact that about two decades have passed since the accelerated expansion of the Universe was discovered \cite{1999ApJ...517..565P} \cite{1998AJ....116.1009R}, there is still no convincing theoretical explanation based on physical foreground and not only phenomenological, the simplest explanation for the accelerating universe is the cosmological constant $\Lambda$. In this sense, information about the dynamics of the expansion by using the deceleration parameter, helps to clarify this behaviour under different models. The deceleration parameter in a general FLRW cosmology obeys to
\begin{equation}
q(z) =-1 + \frac{\left( 1+z \right)}{E(z)} \frac{dE(z)}{dz},
\label{eq5:5.3}
\end{equation}
\noindent that depends explicitly of the cosmological model studied and its matter-energy content through $E(z)$. In general, if $\Omega_X \neq 0$ is sufficiently large (i.e. $\Omega_X >\Omega_m $), then $q(z)<0$ and $\ddot{a}(z)>0$, it corresponds to an accelerated expansion as is shown by observational data, additionally it also indicates a cosmological constant different from zero. If the acceleration is driven by a non perfect fluid, it is important to identify signs to determine if the energy density of the fluid remains constant or dynamic. This is achieved by considering the equation of state (EoS), which, given a cosmological model can be written as \cite{Saini}
\begin{equation}
w (z) = \frac{-1 + \frac{2(1+z)}{3}\frac{dLnH(z)}{dz}}{1-\frac{\Omega_m(1+z)^3}{E^2(z)}}.
\end{equation}
Clearly, $w(z)$ has a dynamical nature given its dependence with redshift, and as mentioned in the introduction, depending on its value the models can be classified as quintessence if $w(z)> -1$ or phantom if $w(z)< -1$. 

In the first group, the accelerating expansion and properties of DE implying a negative pressure ($w(z)< −1/3$), whose simplest example is the cosmological constant ($w(z)=-1$). In the second group, the Einstein's field equations are modified and the new equations combined with the assumption of homogeneity and isotropy lead to a generalized Friedman equation, but $w(z)$ can not be interpreted as a perfect fluid. In this sense, the parameter $w(z)$ determines not only the gravitational properties of DE but also its evolution.

\subsection{$\Lambda$CDM model}

We start the analysis with the standard cosmological model. In this paradigm, the DE is provided by the cosmological constant $\Lambda$,  with an EoS such that $w=-1$ (Figure \ref{fig:wz}). The dimensionless Hubble parameter $E^{2}(z,\Theta)$ is given by
\begin{equation}
E^{2}(z,\Theta) = \Omega_{r}(1+z)^{4} + \Omega_{m}(1+z)^{3} + \Omega_{k}(1+z)^{2} + \Omega_{X},
\label{eq2:2.3}
\end{equation} 
\noindent where $\Omega_{m}$ and  $\Omega_{X} = \Omega_{\Lambda}=1-\Omega_{m}-\Omega_{k}-\Omega_{r}$ are the density parameters for matter and DE respectively and 
 $\Omega_{r}$ corresponds to the radiation parameter. The parameter vector is $\Theta_i^{\Lambda CDM}=\left\lbrace h,\sigma_8, \Omega_{m},\Omega_{k}\right\rbrace$ and the best-fit results are shown in the Table \ref{tab:LCDM}.
\begin{table*}[h]
\centering
\tbl{Summary of best-fit values for $\Lambda CDM$ model.}
{\begin{tabular}{lcc}
\hline \hline 
\textit{Parameter} & \textit{CMB+BAO+G} & \textit{CMB+BAO+G+SNIa+Hz} \\
\hline 
\hline 
h                  & $0.658\pm 0.022$     & $0.6576\pm 0.0068$ \\ 
$\Omega_m$         & $0.339\pm 0.028$     & $0.3126\pm 0.0081$ \\
$\Omega_k$         & $0.004\pm 0.018$     & $-0.0054\pm 0.0035$ \\ 
$\sigma_8$         & $0.733\pm 0.022$     & $0.744\pm 0.019$ \\ 
$\chi_{min}^{2}$   & 28.894               & 621.624 \\
\hline \hline 
\end{tabular}\label{tab:LCDM}}
\end{table*}

\begin{figure}
\centering
   \centerline{\psfig{file=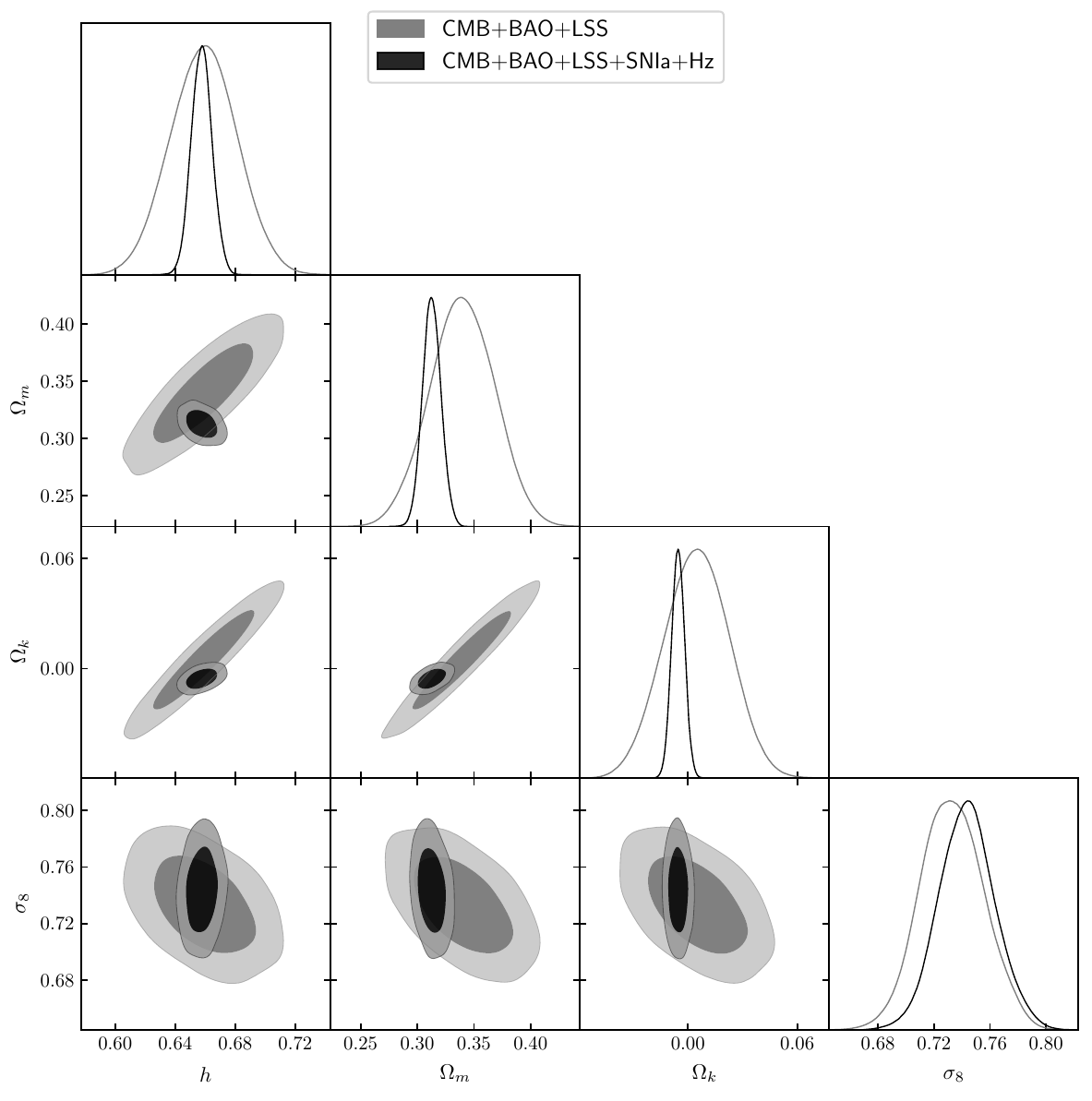,width=\textwidth}}  
   \caption{2D contour plots at $1\sigma$ and $2\sigma$ levels and 1D posterior distributions, with CMB+BAO+G (Black) and CMB+BAO+SNIa+G+Hz (Gray) for $\Lambda CDM$ model.}
    \label{fig:LCDM}
\end{figure}

Note that $\sigma_8$ is a free parameter in all cosmological models, corresponding to RMS mass fluctuations, obtained from the growth parameter $A_{obs}(z)$. Using a $\chi^2$ estimation we find $\sigma_8=0.744\pm0.021$ with 68\% confidence level (see Table \ref{tab:LCDM}). This result is compatible with the one obtained by Planck 2018, which reports a lower uncertainty \cite{2018arXiv180706209P}. Additionally the best-fit value for the DE density normalised to the critical density today is $\Omega_\Lambda=0.687\pm0.009$ at 68\%, which agrees with the limits reported by Planck 2018 ($\Omega_\Lambda=0.6847\pm0.0073$ at 68\% using TT,TE,EE+LowE+lensing)\cite{2018arXiv180706209P}. The value of cosmological constant in this case is positive and different from zero ($\Lambda = 1.5168 \pm 0.0092 \times 10^{-35}s^{-2}$). This value of $\Lambda$ is consistent with measurements obtained by the High-Z Supernova Team and the Supernova Cosmology Project \cite{1999ApJ...517..565P} \cite{2001AIPC..586..316C}. Some derived parameters for this model are shown in Table \ref{tab:par2}, while Figure \ref{fig:LCDM} presents the 2D contour plots at $1\sigma$ and $2\sigma$ levels and 1D posterior distributions with CMB+BAO+G (Black) and CMB+BAO+SNIa+G+Hz (Gray). In Table \ref{tab:LCDM} is evident the impact of adding the SNIa and Hz datasets to CMB + BAO + G, which evidently improves the constraints on the parameters.\\


\subsection{wCDM model}

An extension of the standard model where $w=-1$, is obtained by considering the EoS still constant but with a value deviated from $-1$. In this case the dimensionless Hubble parameter $E^{2}(z,\Theta)$ for a universe with curvature reads as
\begin{eqnarray}
E^{2}(z,\Theta) = \Omega_{r}(1+z)^{4} + \Omega_{m}(1+z)^{3} + \Omega_{k}(1+z)^{2} +
\Omega_{X} (1+z)^{3(1+w)},
\label{eq2:2.4}
\end{eqnarray}
\noindent where $\Omega_{X}=1-\Omega_{m}-\Omega_{k}-\Omega_{r}$. In this model the set of free parameters is given by \linebreak $\Theta_i^{\omega CDM}=\left\lbrace h,\sigma_8, \Omega_{k}, \Omega_{m},\omega\right\rbrace$, and the best-fit values are displayed in Table \ref{tab:wCDM}.
\begin{table*}[h]
\tbl{Summary of the best-fit values for wCDM model.}
{\begin{tabular}{lcc}
\hline \hline 
\textit{Parameter} & \textit{CMB+BAO+G} & \textit{CMB+BAO+G+SNIa+Hz} \\
\hline 
\hline 
h                 & $0.606\pm 0.083$  & $0.676\pm 0.011$     \\ 
$\Omega_m$        & $0.341\pm 0.026$  & $0.3054\pm 0.0087$     \\
$\Omega_k$        & $-0.009\pm 0.029$ & $-0.0048\pm 0.0033$  \\ 
w                 & $-0.86\pm 0.21$   & $-1.070\pm 0.036$    \\ 
$\sigma_8$        & $0.747\pm 0.037$  & $0.738\pm 0.020$     \\ 
$\chi_{min}^{2}$  & 39.366            & 629.191              \\
\hline \hline 
\end{tabular}\label{tab:wCDM}}
\end{table*}
\begin{figure}[htb]
\centering
\begin{center}
   \centerline{\psfig{file=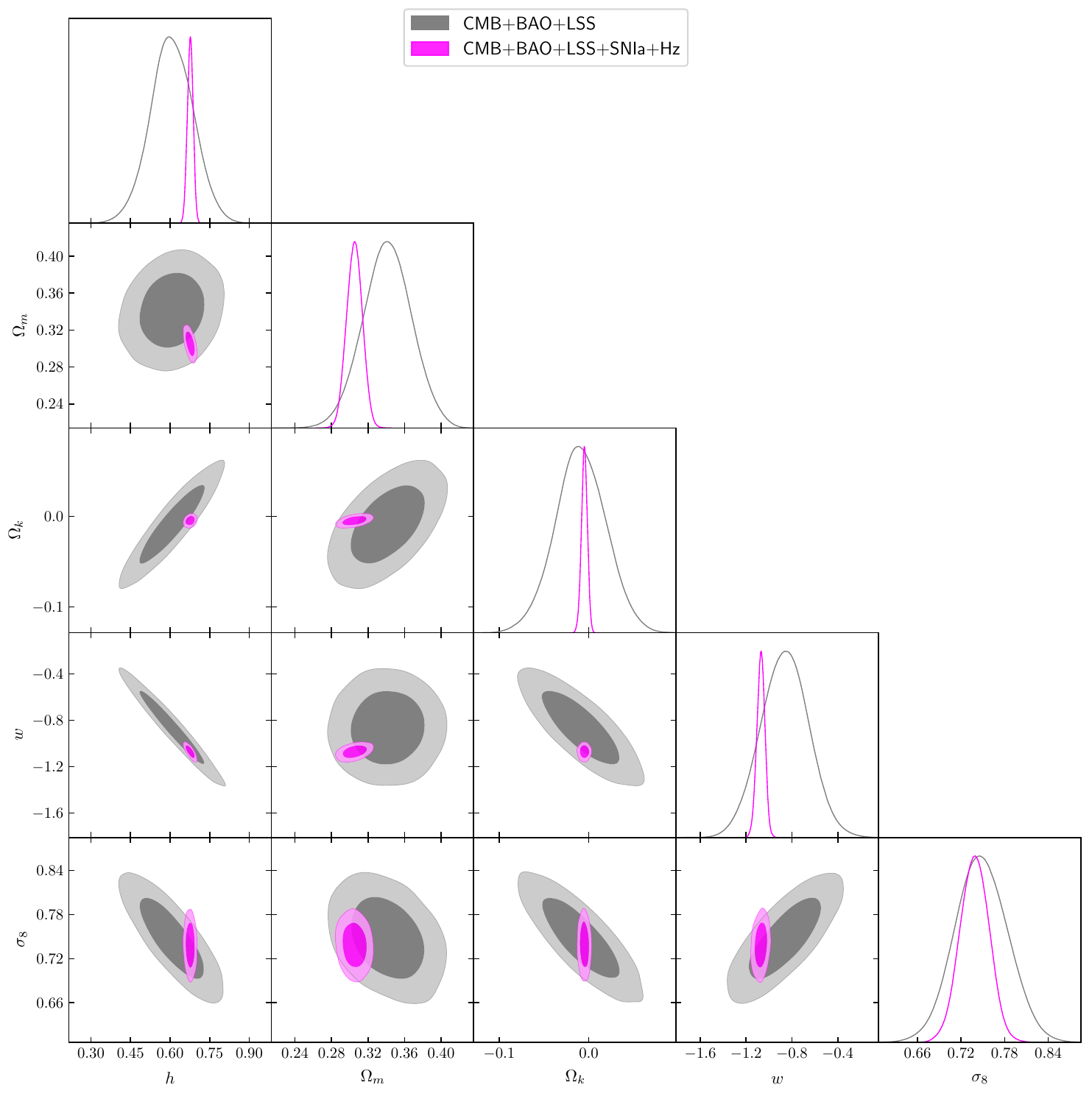,width=\textwidth}}  
   \caption{2D contour plots at $1\sigma$ and $2\sigma$ levels and 1D posterior distributions, with CMB+BAO+G (Magenta) and CMB+BAO+SNIa+G+Hz (Gray) for wCDM model.}
    \label{fig:wCDM}
\end{center}
\end{figure}

Figure \ref{fig:wCDM} shows the 2D contour plots at $1\sigma$ and $2\sigma$ confidence levels and the 1D posterior distribution using CMB+BAO+G (Magenta) and CMB+BAO+SNIa+G+Hz (Gray) for the wCDM model. From this plot we can see that $\Lambda CDM$ model ($\omega=-1$) is still allowed to $1\sigma$ with CMB+BAO+G and combining all datasets, it is consistent with a cosmological constant (see Table \ref{tab:wCDM}). The result obtained in \cite{Shi:2012ma} ($w=-0.990 \pm 0.041$) is also consistent with our results. Recently in \cite{2018arXiv180706209P} the result obtained for the equation of state of the $\omega CDM$ model is  $w_0=-1.03 \pm 0.03$, whose results are consistent with our constraints to $1\sigma$ and $2\sigma$. In this case, from Table \ref{tab:wCDM}, the EoS correspond to a quintessence model for CMB+BAO+G (Fig. \ref{fig:wz}) and phantom when SNIa and Hz are added.\\


\subsection{Chevalier-Polarski-Linder model}
This model corresponds to an extension of the standard scenario considering that the equation of state of DE varies with redshift via the Chevallier-Polarski-Linder (CPL) parametrization \cite{Chevallier:2000qy} \cite{Linder:2003nc} given by
\begin{equation}
w(z) = w_{0} + w_{1} \frac{z}{1+z},
\label{eq2:2.5}
\end{equation}
\noindent where  $w_{0}$ y $w_{1}$ are constants to be fitted. The dimensionless Hubble parameter $E(z)$ for CPL parametrization is written as 
\begin{equation}
 E^{2}(z,\Theta) = \Omega_{r}(1+z)^{4} + \Omega_{k} (1+z)^2+\Omega_{m}(1+z)^{3} + 
 \Omega_{X} X(z),
 \label{eq2:2.6}
\end{equation}
\noindent with $\Omega_{X} =  \left( 1 - \Omega_{k} - \Omega_{m} -\Omega_{r} \right)$ and $X(z)=(1+z)^{3(1 + w_{0} + w_{1})} \exp \left[ - \frac{3w_{1}z}{1+z}\right]$. The set of free parameters constrained are $\Theta_i^{CPL}=\left\lbrace h,\sigma_8,\Omega_{k},\Omega_{m}, w_{0}, w_{1} \right\rbrace$. Table \ref{tab:CPL} shows the best-fit values obtained by using all the observational tests.
\begin{table*}[h]
\tbl{Summary of the best-fit values for CPL model.}
{\begin{tabular}{lcc}
\hline \hline 
\textit{Parameter} & \textit{CMB+BAO+G} & \textit{CMB+BAO+G+SNIa+Hz} \\
\hline 
\hline 
h                 & $0.57\pm 0.11$    & $0.678\pm 0.011$ \\ 
$\Omega_m$        & $0.306\pm 0.079$  & $0.303\pm 0.010$ \\
$\Omega_k$        & $-0.025\pm 0.065$ & $-0.0054\pm 0.0034$ \\ 
$w_{1}$           & $0.57\pm 2.79$    & $0.02\pm 0.53$ \\ 
$w_{0}$           & $-0.92\pm 0.89$   & $-1.091\pm 0.092$     \\ 
$\sigma_8$        & $0.777\pm 0.034$  & $0.735\pm 0.020$ \\ 
$\chi_{min}^{2}$  & 27.353            & 616.376   \\
\hline \hline 
\end{tabular}\label{tab:CPL}}
\end{table*}
\begin{figure}[htb]
\centering
\begin{center}
   \centerline{\psfig{file=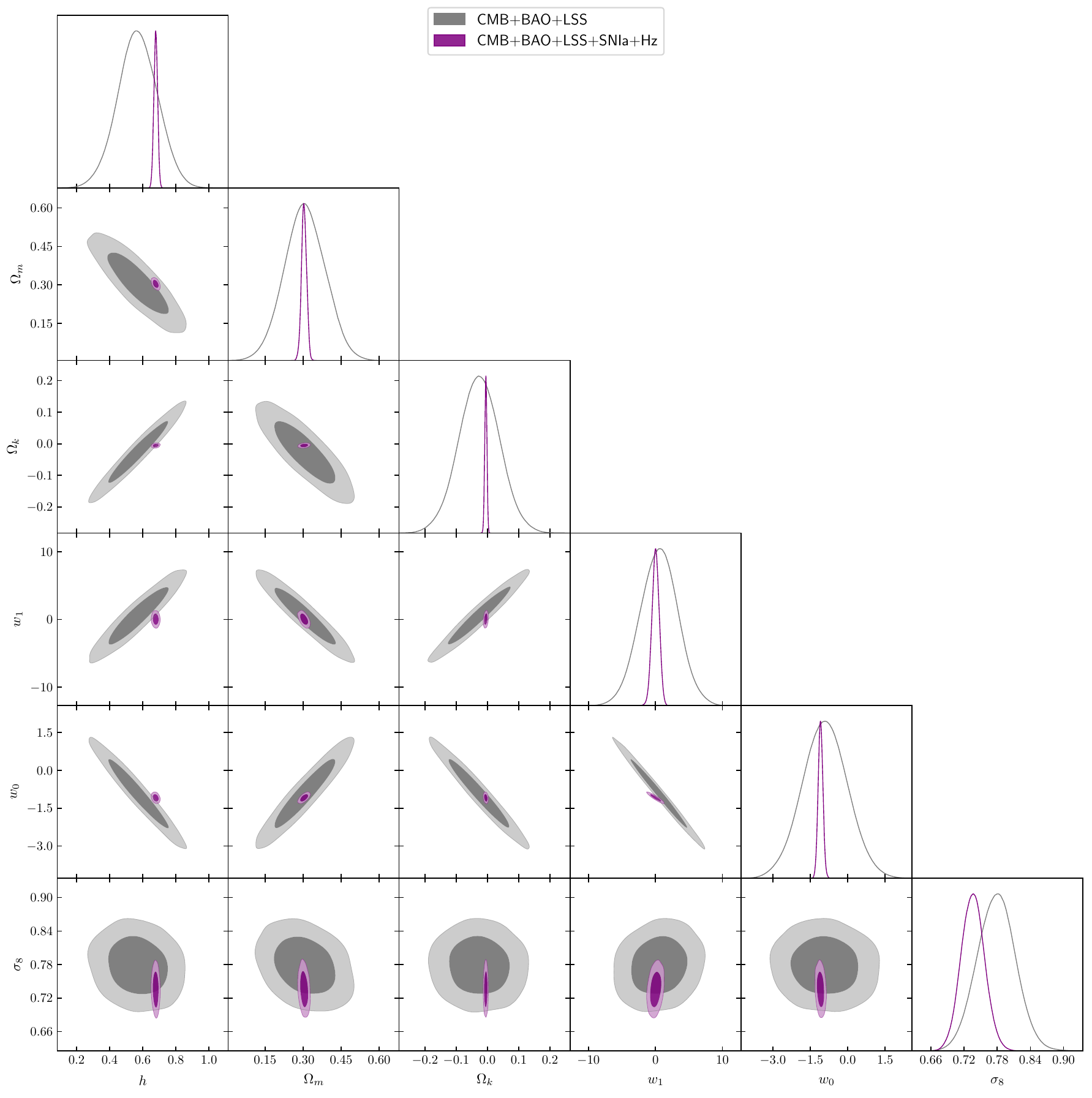,width=\textwidth}}  
   \caption{2D contour plots at $1\sigma$ and $2\sigma$ confidence levels and 1D posterior distributions, with CMB+BAO+G (Purple) and CMB+BAO+SNIa+G+Hz (Gray) for CPL model.}
    \label{fig:CPL}
\end{center}
\end{figure}

Figure \ref{fig:CPL} shows the confidence contour plots at $1\sigma$ and $2\sigma$ levels and the posterior distributions, with CMB+BAO+G (Purple) and CMB+BAO+SNIa+G+Hz (Gray) for the CPL parametrization. The limits on \textit{h}, $\Omega_m$ and $\sigma_8$ using all cosmological data are compatible with the values obtained in \cite{Shi:2012ma} and \cite{2018arXiv180706209P}. However, there is a degeneracy between the curvature parameter $\Omega_k$ and the equation of state $\omega_0$. Recently \cite{2018arXiv180706209P} combine Planck+SNe+BAO datasets getting $\omega_0 = -0.961 \pm 0.077$, which is in good agreement with our estimates. The CPL model is reduced to $\Lambda CDM$ if $\omega_0 = -1$ and $\omega_a = 0$, where is possible to appreciate that the cosmological constant is still allowed in this analysis (see Figure \ref{fig:CPL}). The main physical parameters derived for this model are displayed in Table \ref{tab:par2}, being very close to the reference $\Lambda CDM$ model. From Table \ref{tab:CPL}, the EoS corresponds to a quintessence model for CMB+BAO+G and phantom with the full dataset. On the other hand, in Figure \ref{fig:wz} we can see the evolution of EoS with CMB+BAO+G does not cross the phantom line at late times.\\


\subsection{Interacting Dark Energy model}

In interacting dark energy (EDE) scenarios there is a relation between the energy density of DE $\rho_{x}$ and the density of DM $\rho_{m}$ that could alleviate the cosmic coincidence problem. A general approach is to introduce an interacting term in the right side of continuity equations in the following way \cite{Amendola:1999er,CalderaCabral:2008bx,Cai:2004dk,Dalal:2001dt,Guo:2007zk}
\begin{eqnarray}
\dot{\rho}_{m} &+& 3H\rho_{m} =\delta H \rho_{m},\nonumber\\
\dot{\rho}_{x} &+& 3H\left(1 + w_{x}\right)\rho_{x}= -\delta H \rho_{m},
\label{eq2:2.8}
\end{eqnarray}
\noindent where $w_{x}$ is the equation of state of DE and $\delta$ is an interacting term to be fitted with the observations. Thus, the dimensionless Hubble parameter for this interacting model is described by
\begin{eqnarray}
E^{2}(z,\Theta) = \Omega_{r}(1+z)^{4} + \Omega_{k}(1+z)^{2} +  \Omega_{m}\Psi (z) +  
 \Omega_{X} (1+z)^{3(1 + w_{x})},
\label{eq2:2.9}
\end{eqnarray}
\noindent with $\Omega_{X} = (1 - \Omega_{m} - \Omega_{k}-\Omega_{r})$ and
\begin{equation}
\Psi (z) = \frac{\left( \delta(1+z)^{3(1 + w_{x})} +  3 w_{x}(1+z)^{3 - \delta} \right)}{\delta + 3 w_{x}}.
\label{eq2:2.10}
\end{equation}
\noindent This model is characterised by six parameters $\Theta_i^{IDE}=\left\lbrace h,\sigma_8, \Omega_{k},\Omega_{m}, w_{x},\delta \right\rbrace$, their best-fit values are shown in Table \ref{tab:IDE}.
\begin{table*}[h]
\tbl{Summary of the best-fit values for IDE model.}
{\begin{tabular}{lcc}
\hline \hline 
\textit{Parameter} & \textit{CMB+BAO+G} & \textit{CMB+BAO+G+SNIa+Hz} \\
\hline 
\hline 
h                 & $0.87\pm 0.46$    & $0.688\pm 0.012$ \\ 
$\Omega_m$        & $0.317\pm 0.026$  & $0.276\pm 0.014$ \\
$\Omega_k$        & $0.024\pm 0.046$  & $-0.0183\pm 0.0068$ \\ 
$w_{x}$           & $-1.090\pm 0.38$  & $-0.976\pm 0.057$ \\ 
$\delta$          & $-0.020\pm 0.015$ & $-0.0192\pm 0.0093$     \\ 
$\sigma_8$        & $0.737\pm 0.045$  & $0.769\pm 0.027$ \\ 
$\chi_{min}^{2}$  & 28.669            & 612.756 \\
\hline \hline 
\end{tabular}\label{tab:IDE}}
\end{table*}

\begin{figure}[htb]
\centering
\begin{center}
   \centerline{\psfig{file=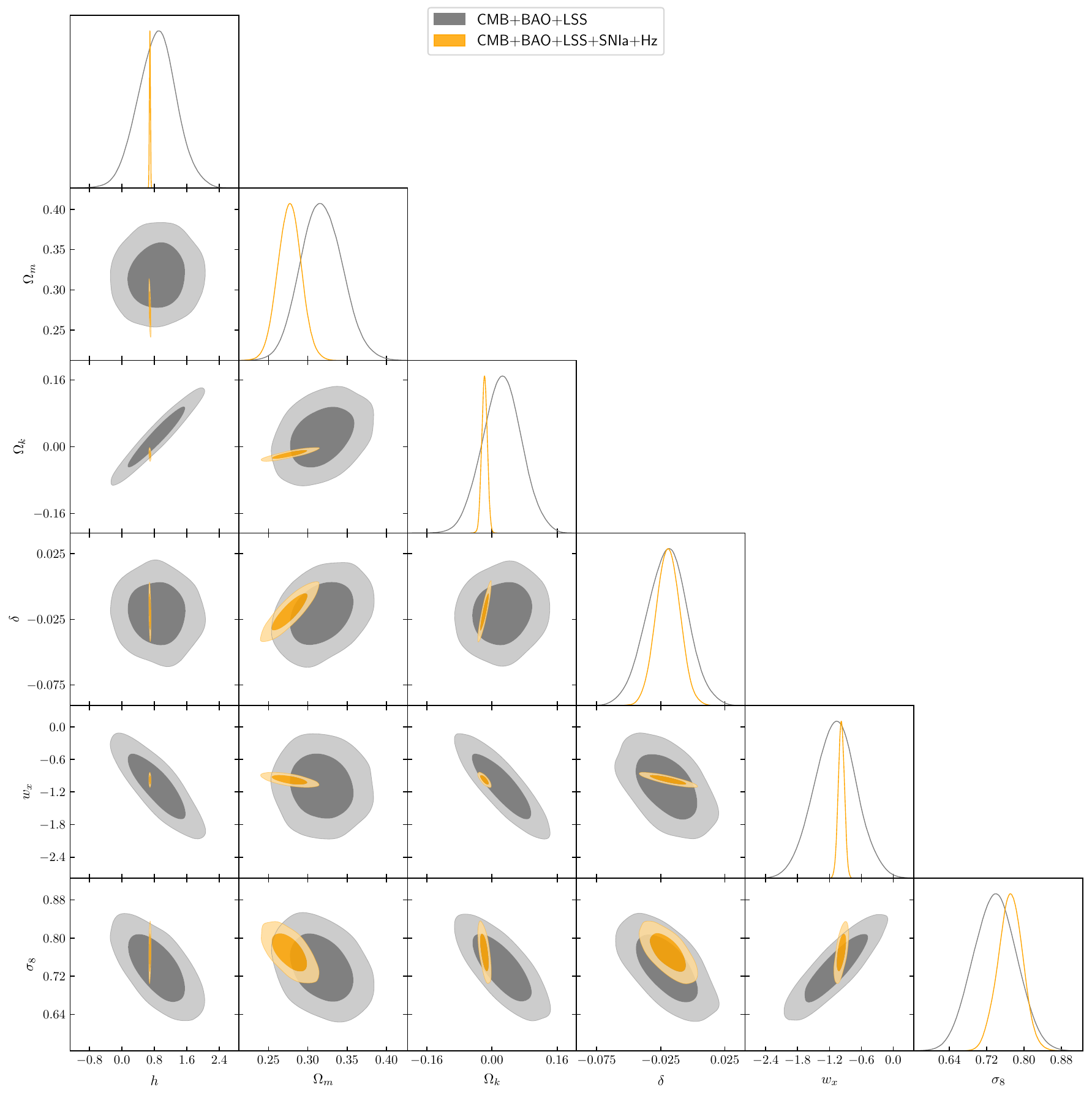,width=\textwidth}}  
   \caption{2D contour plots at $1\sigma$ and $2\sigma$ confidence levels and 1D posterior distributions, with CMB+BAO+G (Orange) and CMB+BAO+SNIa+G+Hz (Gray) for IDE model.}
    \label{fig:IDE}
\end{center}
\end{figure}

Figure \ref{fig:IDE} shows the contour plots at $1\sigma$ and $2\sigma$ confidence level and the posterior distributions for the IDE model using data from CMB+BAO+G (Orange) and from CMB+BAO+SNIa+G+Hz (Gray). In this case, the $\Lambda CDM$ scenario is recovered if $w _x = -1$ and $\delta = 0$. However, the results show that the $\Lambda CDM$ model is ruled out at least $1\sigma$ level in our analysis with all datasets (See Table \ref{tab:IDE}). On the other hand, if the coupling term in equation (\ref{eq2:2.8}) takes a negative value ($\delta < 0$) then there is a transfer from DM to DE, whereas a positive coupling term ($\delta > 0$) implies the opposite, and we can see that the transfer from DM to DE is favoured in this work. The present analysis of the EoS of DE shows a phantom behaviour ($ w_x <$ -1) with CMB+BAO+G dataset, which is consistent with the values obtained by \cite{Shi:2012ma}. In Table \ref{tab:IDE} we can see that the EoS ($w_x$) is phantom by considering only CMB+BAO+G and quintessence with CMB+BAO+G+SNIa+Hz. Figure \ref{fig:wz} shows the evolution of EoS with CMB+BAO+G, which crosses the phantom line at $z\sim 0.64$ from of a quintessential behaviour at early times to a phantom behaviour at late times.\\


\subsection{Early Dark Energy model}

In early dark energy (EDE) scenarios, the energy density of DE is assumed to be significant at high redshifts. This can be possible if the DE tracks the dynamics of the background fluid density \cite{SteinhardtEDE, Wetterich:1988}, especially, this feature could ameliorate the coincidence problem of the cosmological constant. We adopt a general EDE model proposed by \cite{2006JCAP...06..026D} adding a curvature term, what leads to the dimensionless Hubble parameter
\begin{equation}
E^{2}(z,\Theta) = \frac{\Omega_{r}(1+z)^{4} +\Omega_{m}(1+z)^{3} + \Omega_{k}(1+z)^{2}}{1 - \Omega_{X}},
\label{eq2:2.11}
\end{equation}
\noindent with $\Omega_{X}$ given by
\begin{eqnarray}
\Omega_{X} = \frac{\Omega_{X_0} - \Omega_{e}\left[ 1 - (1+z)^{3 w_{0}} \right]  }{\Omega_{X_0} + f(z)} + \Omega_{e} \left[ 1 - (1+z)^{3 w_{0}} \right]
\label{eq2:2.12}
\end{eqnarray}
and
\begin{equation}
f(z) = \Omega_{m}(1+z)^{-3w_{0}} + \Omega_{r}(1+z)^{-3w_{0} +1 } + \Omega_{k}(1+z)^{-3w_{0} -1 },
\label{eq2:2.13}
\end{equation}
\noindent such that $\Omega_{X_0} = 1 - \Omega_{m} - \Omega_{k}- \Omega_{r}$ is the current DE density. This model have five free parameters $\Theta_i^{EDE}=\left\lbrace h,\sigma_8,\Omega_{k},\Omega_{m}, \Omega_{e},\omega_{0}\right\rbrace$, whose best-fit values are shown in Table \ref{tab:EDE}.
\begin{table*}[h]
\tbl{Summary of the best-fit values for EDE model.}
{\begin{tabular}{lcc}
\hline \hline 
\textit{Parameter} & \textit{CMB+BAO+G} & \textit{CMB+BAO+G+SNIa+Hz} \\
\hline 
\hline 
h                 & $0.359\pm 0.029$    & $0.683\pm 0.011$ \\ 
$\Omega_m$        & $0.338\pm 0.026$    & $0.275\pm 0.018$ \\
$\Omega_k$        & $-0.117\pm 0.012$   & $-0.0144\pm 0.0064$ \\ 
$w_{0}$           & $-0.5587\pm 0.0091$ & $-1.039\pm 0.043$ \\ 
$\Omega_e$        & $-0.44\pm 0.10$     & $0.061\pm 0.037$     \\ 
$\sigma_8$        & $0.752\pm 0.033$    & $0.771\pm 0.030$ \\ 
$\chi_{min}^{2}$  & 30.912              & 613.638 \\
\hline \hline 
\end{tabular}\label{tab:EDE}}
\end{table*}
\begin{figure}[htb]
\centering
\begin{center}
   \centerline{\psfig{file=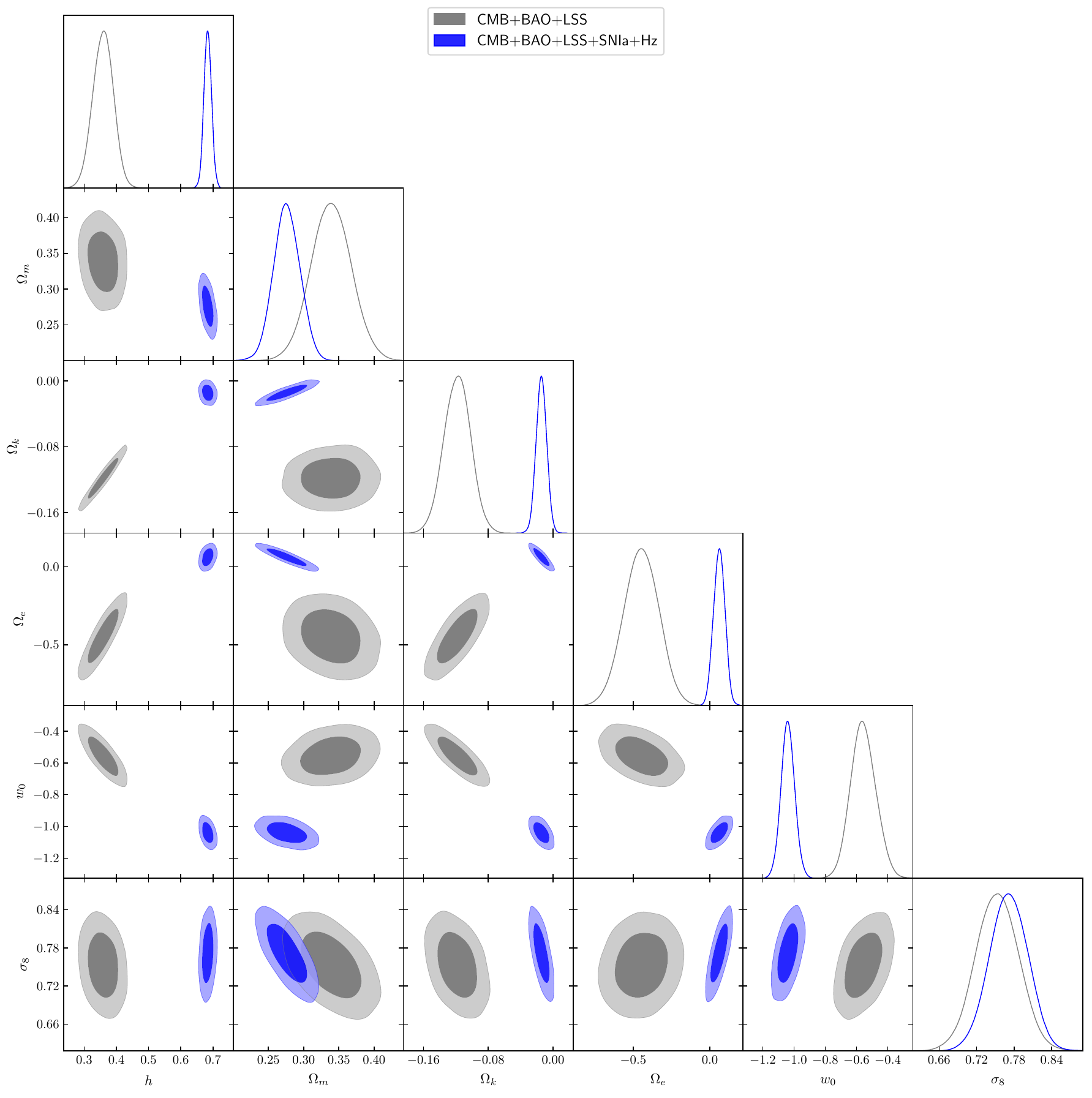,width=\textwidth}}  
   \caption{2D contour plots at $1\sigma$ and $2\sigma$ levels and the posterior distributions, with CMB+BAO+G (Blue) and CMB+BAO+SNIa+G+Hz (Gray) for EDE model.}
    \label{fig:EDE}
\end{center}
\end{figure}

\begin{figure*}[h]
\centering
\begin{center}
 \centerline{\psfig{file=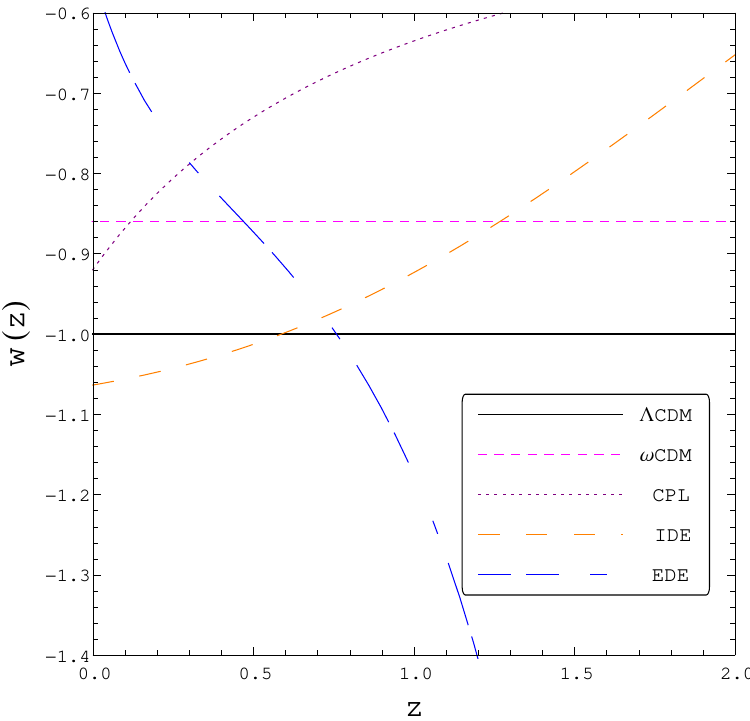,width=12cm}}
    \caption{Evolution of the equation of state $w(z)$ as a function of redshift for $\Lambda CDM$, wCDM, CPL, IDE ($z\sim 0.57$) and EDE ($z \sim 0.77$) models, where the redshift in parenthesis correspond to transition in the phantom line. The color lines correspond to the prediction for each model using the best-fit values from the BAO+CMB+G dataset.}
    \label{fig:wz}
\end{center}
\end{figure*}

After applied the Bayesian analysis we found $\Omega_e=0.061\pm0.037$ with all dataset (see Table \ref{tab:EDE}), which is in accordance at $2\sigma$ confidence level with \cite{2006JCAP...06..026D} who reports a value $\Omega_e<0.04$ at 95\% confidence level by using a combination of data from WMAP+VSA+CBI+BOOMERANG+SDSS+SNIa and it is also in agreement with \cite{2013PhRvD..87h3009P}, reporting a constrain of $\Omega_e<0.015$ at 95\% (see Fig. \ref{fig:EDE}). From the results shown in Figure \ref{fig:EDE}, the $\Lambda CDM$ model ($\Omega_e=0$, $\omega_0 =-1$) is favoured at least $2\sigma$ confidence level with all dataset and discarded for combination CMB+BAO+G. In addition to this, some tensions are found between $h$ and the rest of parameters, in particular significant deviations when SNIa+Hz are included appears in the limits of the posterior distribution. Table \ref{tab:EDE} shows that the EoS ($w_0$) is quintessence for CMB+BAO+G and phantom by adding SNIa+Hz, on the other hand, the Figure \ref{fig:wz} displays the evolution of EoS with CMB+BAO+G which crosses the phantom line at $z\sim 0.49$ from a phantom behaviour at early times to a quintessential behaviour at late times.\\

\subsection{Exclusion analysis}

By comparing the absolute and relative differences obtained after 
computing the exclusion criteria AIC/BIC, we discern the most favoured 
model in terms of its best-fit values and the number of parameters used in the Bayesian analysis. Table \ref{tab:AIC/BIC} shows the values of $\Delta AIC$ and $\Delta BIC$ for DE models from all cosmological tests. The IDE model gives the lowest value of $\Delta AIC$ and $\Delta BIC$, therefore, we conclude this is the model most favoured by observational data, as it can also analysed in Table \ref{tab:par2}. The $\Delta AIC$ and $\Delta BIC$ values for the other models are measured with respect to IDE. Following \cite{Shi:2012ma}, the DE models can be classified in two groups: {\emph 1)} models with models that show a substantial level of empirical support to IDE, EDE and positive evidence for $\Lambda CDM$; {\emph 2)} models with a considerably low level of empirical support and positive evidence against to CPL and $\omega CDM$ models.

\begin{table}[h]
\tbl{Comparison of the different cosmological models with the 
$\Delta AIC$ y $\Delta BIC$ criteria using combined analysis dataset (CMB+BAO+G+SNIa+H(z)), where N=639 and $\chi^2_{red}=\chi^2_{min}/\nu$, where $\nu$ is the degrees of freedom usually given by N - k.}
{\begin{tabular}{lccccccc}
\hline
 Model & k  & $\chi_{red}^{2}$ & $AIC$ &  $\Delta AIC$  & $BIC$ &  $\Delta BIC$ \\
\hline
\hline
$\Lambda$ CDM   & 4 & 0.980 & 631.624 & 6.868 & 653.924 & 2.409  \\
$\omega$ CDM    & 5 & 0.990 & 637.191 & 12.435& 655.031 & 3.516  \\
CPL             & 6 & 0.973 & 628.376 & 3.620 & 655.135 & 3.620  \\ 
IDE             & 6 & 0.968 & 624.756 & 0.000 & 651.515 & 0.000   \\
EDE             & 6 & 0.969 & 625.638 & 0.882 & 652.397 & 0.882  \\
\hline
\end{tabular}\label{tab:AIC/BIC}}
\end{table}

\subsection{History of the expansion and growth of structures}

\begin{figure*}[h]
\centering
\begin{center}
 \centerline{\psfig{file=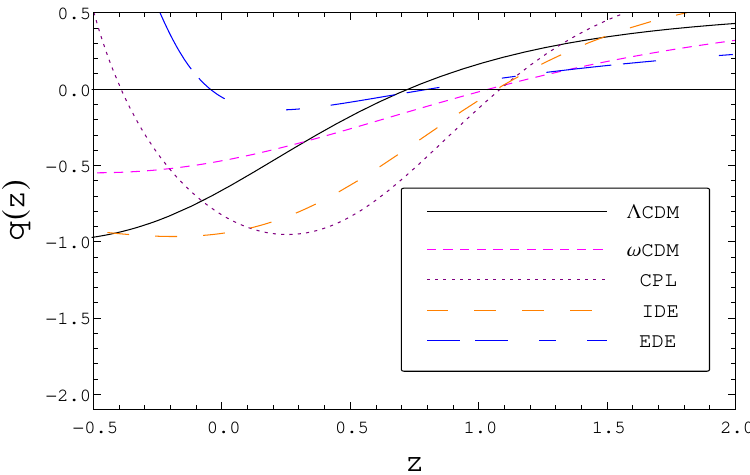, width=\textwidth}}
    \caption{Deceleration parameter as a function of redshift using BAO and G dataset. The transition from decelerated to accelerated range is shown ($q(z_t)=0$), and the current value of ($q_0$) ($\Lambda CDM$ ($z_t\sim 0.73$, $q_0=-0.67$), $\omega CDM$ ($z_t\sim 1.07$, $q_0=-0.50$), CPL ($z_t\sim 1.07$, $q_0=-0.79$), IDE ($z_t\sim 1.07$,  $q_0=-0.95$), EDE ($z_t\sim 0.84$, $q_0=-0.06$)).  Note the peculiar behaviour of the deceleration parameter to later times for dynamical DE models (CPL, IDE, EDE).}
    \label{fig:qz}
\end{center}
\end{figure*}

Figure \ref{fig:qz} displays the behaviour of the deceleration parameter $q(z)$ as a function of redshift using only data from BAO and Growth factor dataset (see Table \ref{tab:BAOqz}). As expected, the models studied have $q(z)<0$ at late times and  $q(z)>0$ at earlier epoch, it means that the history of the expansion is slowed down in the past and accelerated today. All cosmological models present a transition (at $z_t$) between the two periods, however, models with dynamical DE have an interesting behaviour of slowing down of acceleration at low redshift, i.e., late times, using only data from BAO+G, which can be characterised through the change of sign in the jerk parameter $j(z)$ \cite{2018Univ....4...21B} (CPL: $j(z_{low}) \rightarrow 0$, when $z_{low} \sim 0.25$; IDE: $j(z_{low})\rightarrow 0$, when $z_{low} \sim -0.09$; EDE: $j(z_{low}) \rightarrow 0$, when $z_{low} \sim 0.24$). The $j(z)$ parameter can be interpreted as the slope at each point of $q(z)$ that indicates a change of acceleration. This result is consistent with the one presented by J. Barrow, R. Bean and J. Magueijo \cite{2000MNRAS.316L..41B}, who raises the possibility of a scenario consistent with the current accelerating universe and does not involve an eternal accelerated expansion. In \cite{2012PhyU...55A...2B}, an extensive analysis is made exploring this possibility. This can be also a clear behaviour of a dynamical DE at low redshift in these models with variation of the density of DE over time.

\begin{table*}[h]
\tbl{Best-fit values obtained from the Bayesian analysis using only BAO and Growth dataset for each model considered in this work.}
{\begin{tabular}{lcl}
\hline
 Model & $\chi_{min}^{2}$ & Parameters \\
\hline
\hline
$\Lambda$CDM & 25.71 & h=0.7497, $\Omega_m$=0.3202, $\Omega_k$=-0.1462   \\
$\omega$CDM   & 15.29 & h=0.6133, $\Omega_m$=0.3067, $\Omega_k$=-0.6021,  $\omega$=-0.6548\\
CPL                     & 13.62 & h=1.9489, $\Omega_m$=0.2829, $\Omega_k$=-0.4470,  $\omega_a$=0.8145, $\omega_0$=-0.8927\\ 
IDE                      & 15.29 & h=0.6245, $\Omega_m$=0.3064, $\Omega_k$=-0.6024, $\omega_x$=-0.6554, $\delta$=-0.0027 \\
EDE                     & 15.43 & h=0.3885, $\Omega_m$=0.3419, $\Omega_k$=-0.2887, $\omega_0$=-0.4582, $\Omega_e$= -0.5558\\
\hline
\end{tabular}\label{tab:BAOqz}}
\end{table*}

\begin{figure*}[h]
\centering
\begin{center}
   \centerline{\psfig{file=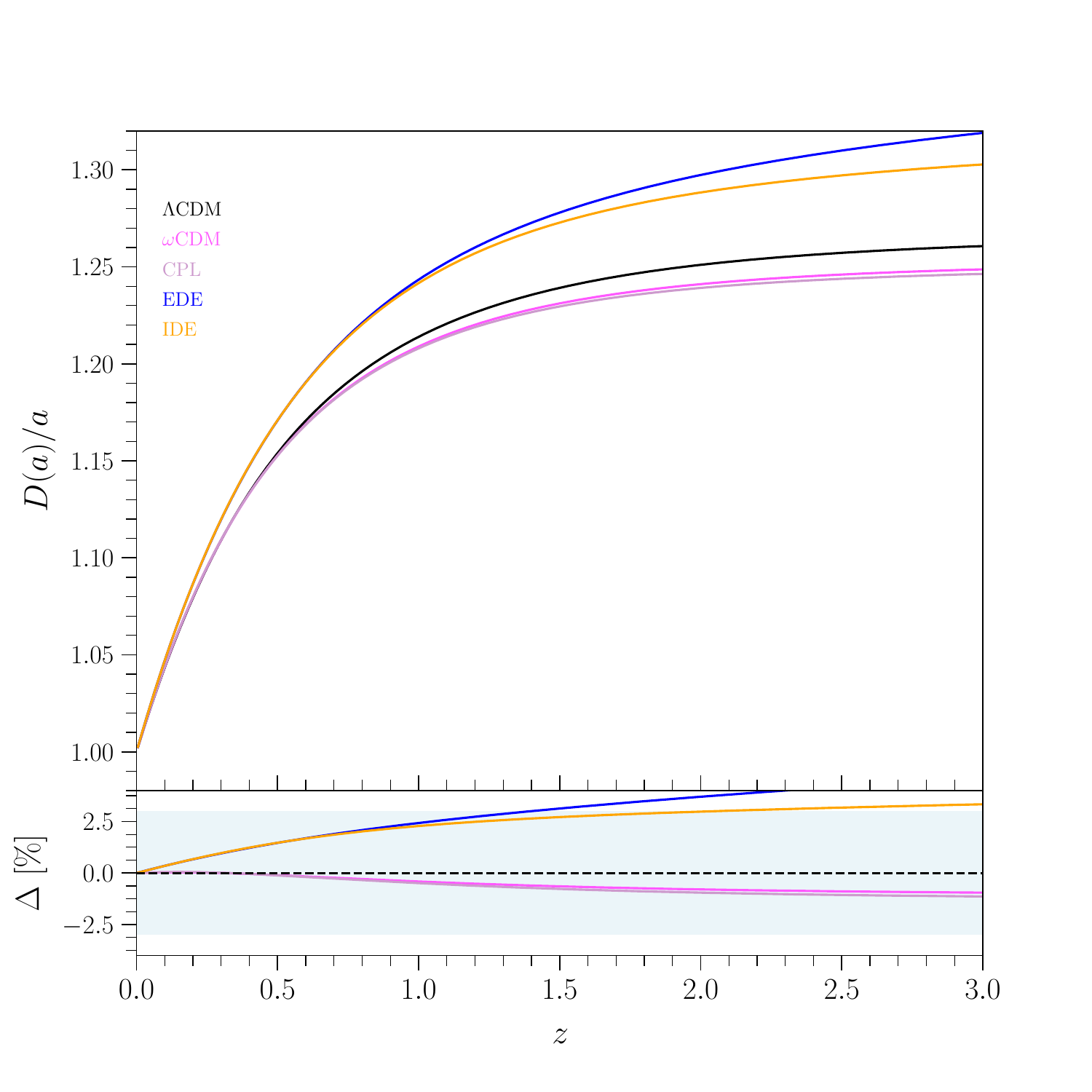,width=12cm,height=9.5cm}}
    \caption{Normalised growth factor ratio $D(a)/a$ as a function of redshift. The shaded region in the lower panel represents a $3\%$ deviation around the $\Lambda CDM$ model prediction. Best-fit parameters from CMB+BAO+G+SNIa+Hz constraints have been use for each model.}
    \label{fig:GrowthFactor}
\end{center}
\end{figure*}

By assuming general relativity as the correct theory of gravity, we use measures of the growth parameter $A(z)$ (See Table \ref{tab:Az} in Appendix \ref{AppxC}) to constrain independently the mass variance of fluctuations $\sigma _8$. This method allows to break the $\Omega _m-\sigma _8$ degeneracy, through the use of equation (\ref{eq3:3.24}) to thereby achieve a good independent constrain. Figure \ref{fig:GrowthFactor} shows the evolution of the normalised growth factor computed by using equation \ref{eq3:3.24} for each model considered in our research. Deviations around 3\% with respect to $\Lambda$CDM can be appreciated in the lower panel, where all models are in well agreement at low redshift. In particular, the deviations increase above 1\% from redshift 0.5 at all redshifts considered, where the transition to an accelerated stage occurs, i.e.,  $q(z)<0$. As the growth factor evolves as a function of redshift, the $w$CDM and CPL models remain close to $\Lambda$CDM.

\section{Summary and discussion}

In this paper we perform a dynamical analysis of different dark energy models, including a comparison between them through the best-fit to observational data using the most recent information from CMB and LSS.\\

We studied the history of the of cosmic expansion through the $q(z)$ parameter with data from LSS, using BAO distance ratio scale $r_{s}/D_v (z)$ and the growth factor. We found new evidence on some results indicated in previous works \cite{2018Univ....4...21B}, showing a peculiar behaviour of the deceleration parameter $q(z)$ at late times ($z_{low}< 0.5$), that is indicated by the change of sign in the jerk parameter $j(z)$ (+$\rightarrow$ -), as consequence the Universe could pass to an decelerate stage in the near future (Figure \ref{fig:qz}). This phenomenon raises the possibility that an accelerated expansion does not imply an eternal accelerated expansion, even in presence of DE. This particular behaviour is present only in models with DE density varying with time, and possibly due to the dynamics of the DE density which in principle can be a sign  to distinguish it from a cosmological constant.\\

By using these datasets, we obtained the best-fit parameters and we 
classify the models following the information criteria by  $\Delta AIC$ and the $\Delta BIC$, to compare different results and to see which one is the most favoured by the data employed. Our analysis shows that IDE model is preferred by the Bayesian and Akaike criterion, but although $\Lambda CDM$, $\omega CDM$, EDE and CPL models are less favoured, they are not discarded. This result is very interesting, since models that include interactions in the dark sector are gaining attention in the community of cosmologists because they offer and alternative way to the standard scenario to solve different tensions such as those of $H_0$, as it has been recently studied in \cite{Kumar}.

\section*{Acknowledgments}

The authors thank \textit{Dr. Florian Beutler} for his appropriate comments about the most recent BAO data and \textit{Dr. Santiago Vargas} for his cordial review on systematic errors during the preparation of this paper. A. Bonilla expresses his gratitude to Universidad Distrital FJDC for the academic support and funding.

\section{Appendix A}\label{AppxA}

\subsection{Supernova data}

We use the Union $2.1$ compilation, which contains a sample of 580 data points of SNIa. The luminosity distance is obtained through the relation $d_L(z)=(1+z)^2d_A(z)$, and it is fitted to a cosmological model by minimizing the $\chi^2$ function defined by
\begin{equation}
\chi_{SNIa}^2 = \textsf{A} - \frac{\textsf{B}^2}{\textsf{C}}
\end{equation}
\noindent where
\begin{eqnarray*}
\textsf{A} = \sum_{i=1}^{580}\frac{
[\mu_{th}(z_{i},p_i)-\mu_{obs}(z_{i})]^2 }{\sigma_{\mu_i}^2},\\
\end{eqnarray*}
\begin{equation}
\textsf{B} = \sum_{i=1}^{580}\frac{
\mu_{th}(z_{i},p_i)-\mu_{obs}(z_{i}) }{\sigma_{\mu_i}^2},
\end{equation}
\begin{eqnarray*}
\textsf{C} = \sum_{i=1}^{580}\frac{1}{\sigma_{\mu_i}^2},
\end{eqnarray*}
\noindent with  $\mu(z)\equiv 5\log_{10}[d_L(z)/\texttt{Mpc}]+25$ being the theoretical expectation of the distance modulus, and we have marginalized over the nuisance parameters $\mu_0$ and $\mu_{obs}$.

\subsection{Observational Hubble Data}

The differential evolution of early type passive galaxies provides direct information about the Hubble parameter $H(z)$. We adopt
$36$ Observational Hubble Data (OHD) at different redshifts ($0.0708 \leq z \leq 2.36$) obtained from \cite{2015arXiv150702517M}, where $26$ data are deduced from the differential age method, and the remaining $10$ data belong to the radial BAO method. Here, we use these data to constrain the free parameters of the models under consideration. The corresponding $\chi^2$ is defined as
\begin{equation}
\chi_{H(z)}^2 (H_0,p_i) =\sum_{i=1}^{36}\frac{ [H_{obs} (z_i) - H_{th}(z_i,H_0,p_i)]^2 }{\sigma_{H}^2(z_i)},
\end{equation}
\noindent where $H_{th}(z_i,H_0,p_i)$ is the theoretical value of
the Hubble parameter at redshift $z_i$. This equation can be re-written as
\begin{equation}
\chi_{H(z)}^2 (H_0,p_i) = \textsf{A}_1 - \textsf{B}_1 +
\textsf{C}_1,
\end{equation}
\noindent with
\begin{eqnarray}
&&\textsf{A}_1 = H_0^2 \sum_{i=1}^{36}\frac{
E^2(z_i,p_i)}{\sigma_i^2},\nonumber\\ &&\textsf{B}_1 =
2H_0\sum_{i=1}^{36} \frac{H_{obs} (z_i) E^2(z_i,p_i)}{\sigma_i^2},\nonumber\\
&&\textsf{C}_1 = \frac{H^2_{obs} (z_i) }{\sigma_i^2}.
\end{eqnarray}

To marginalize over $H_0$, we assume a Gaussian prior distribution with standard deviation width $\sigma_{H_0}$ and mean $\bar{H}_0$. Then, we build the posterior likelihood function $\mathcal{L}_H(p)$ that depends just on the free parameters $p_i$, as
\begin{equation}
\mathcal{L}_H(p_i) = \int \pi_H (H_0) exp \left[-\chi^2_H(H_0,p_i) \right] dH_0,
\end{equation}
\noindent where
\begin{equation}
\pi_H (H_0) = \frac{1}{\sqrt{2\pi}\sigma_{H_0}}  exp \left[ -\frac{1}{2} \left( \frac{H_0 - \bar{H}_0 }{\sigma_{H_0}}  \right)^2 \right] ,
\end{equation}

\noindent is a prior probability function widely used in the
literature. Finally, we minimize $\chi_{H(z)}^2 (p_i)=-2 \ln
\mathcal{L}_H(p_i)$ with respect to the free parameters $p_i$ to
obtain the best-fit.

\section{Appendix B}\label{AppxB}

Table \ref{tab:par2} displays the main results of derived cosmological parameters from the free parameters constrained in this work considering the full observational data samples.

\begin{table*}[h]
\tbl{Derived parameters for different cosmological DE models. We assume $\Omega_{b_0}=0.045$ \cite{Kirkman:2003uv} and  $N_{eff}=3.04$ \cite{2018arXiv180706209P} for all cosmological models.}
{\begin{tabular}{lccccc}
\hline 
\hline 
Parameter & $\Lambda CDM$  & $\omega CDM$ & \textit{CPL} & \textit{IDE} &  \textit{EDE} \\\hline
$H_0$     & $65.76 \pm 0.68$ & $67.6 \pm 1.1$ & $67.8 \pm 1.1$ & $68.8 \pm 1.2$& $68.3 \pm 1.1$\\
$t_0$     & $14.879 \pm 0.154$ & $14.474 \pm 0.235$ & $14.432 \pm 0.234$ & $14.222 \pm 0.248$ & $14.326 \pm 0.231$  \\
$10^{-5}\Omega_{r_0} $  & $9.66 \pm 0.20$   & $9.140\pm 0.297 $ & $9.086 \pm 0.294$   & $8.824 \pm 0.307$   & $8.953 \pm 0.288$   \\
$10^{-5}\Omega_{\gamma_0} $ & $5.709 \pm 0.181$ & $5.403 \pm 0.176$              & $5.371 \pm 0.174$  & $5.216 \pm 0.182$   & $5.292 \pm 0.170$   \\
$10^{-5}\Omega_{\nu_0}$     & $3.950 \pm 0.082$  & $3.737 \pm 0.122$              & $3.715 \pm 0.120$   & $3.608 \pm 0.125$   & $3.661 \pm 0.117$   \\
$\omega_{m_0}$               & $0.1351 \pm 0.0044$ & $0.140 \pm 0.006$              & $0.139 \pm 0.006$   & $0.131 \pm 0.008$   & $0.128 \pm 0.009$   \\
$\omega_{b_0}$               & $0.0211 \pm 0.0004$ & $0.0223 \pm 0.0007$            & $0.0224 \pm 0.0007$  & $0.0230 \pm 0.0008$ & $0.0227 \pm 0.0007$ \\
$\Omega_{X_0} $              & $0.693 \pm 0.011$   & $0.699 \pm 0.012$              & $0.702 \pm 0.013$   & $0.742 \pm 0.021$   & $0.739 \pm 0.024$   \\
$10^{-30}\rho_{cri_0}$       & $8.130 \pm 0.168$   & $8.591 \pm 0.279$              & $8.642 \pm 0.280$   & $8.898 \pm 0.310$   & $8.770 \pm 0.282$   \\
$10^{-30}\rho_{X_0}$        & $5.632 \pm 0.137$   & $6.001 \pm 0.211$              & $6.069 \pm 0.217$   & $6.604 \pm 0.268$   & $6.483 \pm 0.268$   \\
$c_s$                        & $0.452 \pm 0.002$   & $0.447 \pm 0.003$              & $0.447 \pm 0.002$ & $0.444 \pm 0.003$   & $0.446 \pm 0.003$   \\
$z_{drag}$                   & $1017.13 \pm 1.29$    & $1020.26 \pm 2.02$               & $1020.53 \pm 2.01$  & $1021.16 \pm 2.21$  & $1020.19 \pm 2.05$  \\
$r_{drag}$                   & $153.123 \pm 1.482$   & $151.767 \pm 2.044$            & $152.726 \pm 2.136$ & $148.238 \pm 3.771$ & $149.85 \pm 4.14$ \\
$z_{cmb}$                    & $1093.12 \pm 0.53$    & $1091.71 \pm 0.73$               & $1091.5 \pm 0.8$    & $1089.82 \pm 0.89$  & $1090.1 \pm 1.0$    \\ \hline
\hline
\end{tabular}\label{tab:par2}}
\end{table*}

\section{Appendix C}\label{AppxC}

\begin{table*}[htb]
\tbl{Summary of the observed growth rate and references.}
{\begin{tabular}{cccc}
\hline
Index & $z$ &  $A_{obs}(z_i)$& Refs.\\ 
\hline 
\hline
1   &    0.02  & $0.360\pm 0.040$ & \cite{Hud12}\\
2   &    0.067 & $0.423\pm 0.055$ & \cite{Beutler}\\
3   &    0.17  & $0.510\pm 0.060$ & \cite{Perc04,Song09}\\
4   &    0.18  & $0.360\pm 0.090$ & \cite{Blake}\\
5   &    0.25  & $0.351\pm 0.058$ & \cite{Sam11}\\
6   &    0.37  & $0.460\pm 0.038$ & \cite{Sam11}\\
7   &    0.38  & $0.440\pm 0.060$ & \cite{Blake}\\
8   &    0.41  & $0.450\pm 0.040$ & \cite{2011MNRAS.415.2892B}\\
9   &    0.60  & $0.550\pm 0.120$ & \cite{Pezzotta}\\
10  &    0.60  & $0.430\pm 0.040$ & \cite{2011MNRAS.415.2892B}\\
11  &    0.78  & $0.380\pm 0.040$ & \cite{2011MNRAS.415.2892B}\\
12  &    0.57  & $0.427\pm 0.066$ & \cite{Reid12}\\
13  &    0.30  & $0.407\pm 0.055$ & \cite{Tojeiro:2012rp}\\
14  &    0.40  & $0.419\pm 0.041$ & \cite{Tojeiro:2012rp}\\
15  &    0.50  & $0.427\pm 0.043$ & \cite{Tojeiro:2012rp}\\
16  &    0.60  & $0.433\pm 0.067$ & \cite{Tojeiro:2012rp}\\
17  &    0.86  & $0.400\pm 0.110$ & \cite{Pezzotta}\\
18  &    1,40  & $0.484\pm 0.116$ & \cite{Okumura}\\
\hline
\end{tabular}\label{tab:Az}}
\end{table*}

\end{document}